\documentclass[letterpaper,twocolumn,10pt]{article}
\usepackage{zhanggroup}

\usepackage{tikz}
\usepackage{amsfonts}
\usepackage{xspace} 
\usepackage{amsmath}
\usepackage{mathrsfs}
\usepackage{amssymb}
\usepackage{amsmath}
\usepackage{amsthm}
\usepackage[sort&compress,numbers]{natbib}
\usepackage{subcaption}
\usepackage{booktabs}
\usepackage{balance}
\usepackage{xcolor}
\hypersetup{
  colorlinks,
  linkcolor={blue!60!green},
  citecolor={green!60!blue},
  urlcolor={orange!60!red}
}

\newtheorem{definition}{Definition}
\usepackage[hang,flushmargin]{footmisc}
\usepackage{titlesec}
\titlespacing*{\section}{0pt}{*1.5}{2pt}
\titlespacing*{\subsection}{0pt}{*1.5}{2pt}
\titlespacing*{\subsubsection}{0pt}{*1}{1pt}
\usepackage[absolute]{textpos}

\newcommand{\mypara}[1]{\smallskip\noindent{\bf {#1}.}\xspace}
\newcommand{\systemName}{\textsc{ML-Doctor}\xspace}
\newcommand{\meminf}{\ensuremath{\mathsf{MemInf}}\xspace}
\newcommand{\modelinv}{\ensuremath{\mathsf{ModInv}}\xspace}
\newcommand{\modelsteal}{\ensuremath{\mathsf{ModSteal}}\xspace}
\newcommand{\attrinf}{\ensuremath{\mathsf{AttrInf}}\xspace}
\newcommand{\aux}{\mathsf{aux}}
\newcommand{\target}{\mathsf{target}}
\newcommand{\xtarget}{\ensuremath{x_{\target}}}
\newcommand{\BM}{\ensuremath{\mathcal{M}^{\mathsf{B}}}\xspace}
\newcommand{\WM}{\ensuremath{\mathcal{M}^{\mathsf{W}}}\xspace}
\newcommand{\ND}{\ensuremath{\mathcal{D}_{\aux}^{\mathsf{N}}}\xspace}
\newcommand{\SD}{\ensuremath{\mathcal{D}_{\aux}^{\mathsf{S}}}\xspace}
\newcommand{\PD}{\ensuremath{\mathcal{D}_{\aux}^{\mathsf{P}}}\xspace}
\newcommand{\model}[2]{\ensuremath{\mathcal{M}_{#1}^{#2}}\xspace}
\newcommand{\dset}[2]{\ensuremath{\mathcal{D}_{#1}^{#2}}\xspace}
\newcommand{\atktu}[3]{\ensuremath{\langle\! {#1},\!{#2},\!{#3}\! \rangle}\xspace}
\newcommand{\tuple}[1]{\ensuremath{\langle #1 \rangle}}
\renewcommand{\Pr}[1]{\ensuremath{\mathsf{Pr}\left[#1\right]}\xspace}
\renewcommand{\AA}{\mathcal{A}\xspace}
\newcommand{\calN}{\mathcal{N}\xspace}
\newcommand{\zcdp}{zCDP\xspace}
\newcommand{\customTableFont}{\fontsize{7pt}{8pt}\selectfont}

\begin{document}

\begin{textblock}{12}(2,1)
\centering
To Appear in the 31st USENIX Security Symposium, August 10–12, 2022.
\end{textblock}

\date{}

\title{\Large \bf \systemName: Holistic Risk Assessment of Inference Attacks\\ Against Machine Learning Models}

\author{
{\rm Yugeng Liu\textsuperscript{1}\thanks{The first two authors made equal contributions.}}\ \ \ \ \
{\rm Rui Wen\textsuperscript{1}\textsuperscript{\textcolor{blue!60!green}{$\ast$}}}\ \ \
{\rm Xinlei He\textsuperscript{1}}\ \ \
{\rm Ahmed Salem\textsuperscript{1}}\ \ \
{\rm Zhikun Zhang\textsuperscript{1}}\ \ \ 
\\
{\rm Michael Backes\textsuperscript{1}}\ \ \
{\rm Emiliano De Cristofaro\textsuperscript{2}}\ \ \
{\rm Mario Fritz\textsuperscript{1}}\ \ \
{\rm Yang Zhang\textsuperscript{1}}\ \ \
\\
\\
\textsuperscript{1}\textit{CISPA Helmholtz Center for Information Security}\ \ \ \textsuperscript{2}\textit{UCL \& Alan Turing Institute}
}

\maketitle

\begin{abstract}

Inference attacks against Machine Learning (ML) models allow adversaries to learn sensitive information about training data, model parameters, etc.
While researchers have studied, in depth, several kinds of attacks, they have done so in isolation.
As a result, we lack a comprehensive picture of the risks caused by the attacks, e.g., the different scenarios they can be applied to, the common factors that influence their performance, the relationship among them, or the effectiveness of possible defenses.
In this paper, we fill this gap by presenting a first-of-its-kind holistic risk assessment of different inference attacks against machine learning models.
We concentrate on four attacks -- namely, membership inference, model inversion, attribute inference, and model stealing -- and establish a threat model taxonomy.

Our extensive experimental evaluation, run on five model architectures and four image datasets, shows that the complexity of the training dataset plays an important role with respect to the attack's performance, while the effectiveness of model stealing and membership inference attacks are negatively correlated.
We also show that defenses like DP-SGD and Knowledge Distillation can only mitigate {\em some} of the inference attacks.
Our analysis relies on a modular re-usable software, \systemName, which enables ML model owners to assess the risks of deploying their models, and equally serves as a benchmark tool for researchers and practitioners.\footnote{See \url{https://github.com/liuyugeng/ML-Doctor}.}

\end{abstract}

\section{Introduction}

Over the last decade, research in Machine Learning (ML), and in particular Deep Learning, has made tremendous progress.
However, the deployment and success of these technologies might be affected by attacks against ML models that prompt serious security and privacy risks.
In particular, inference attacks~\cite{SSSS17,PMSW18,SZHBFB19,FJR15,ZJPWLS20,MSCS19,SS20,TZJRR16,OSF19,SBBFZ20,HJBGZ21} allow adversaries to infer information from a target ML model, e.g., about the training data, the model's parameters, and so on.

In this paper, we focus on four representative attacks: membership inference~\cite{SSSS17}, model inversion~\cite{FJR15}, attribute inference~\cite{MSCS19}, and model stealing~\cite{TZJRR16}.
The first three target a model's \textit{training dataset}, aiming to, respectively, determine whether or not an exact data sample belongs to it, recover (part of) it, or predict properties that are not related to the model's original task.
Model stealing involves reconstructing the target model's (non-public) parameters.
Inference attacks can lead to severe consequences, including violating individuals' privacy, as ML models are often trained on sensitive data or compromising the model owner's intellectual property~\cite{C20}.

Overall, existing inference attacks have been studied under different threat models and experimental settings, albeit in isolation.
This prompts the need for a holistic understanding of the risks caused by these attacks, such as the scenarios different inference attacks can be applied to, the common factors that influence these attacks' performance, and the relations among the attacks, as well as the overall effectiveness of defense mechanisms.
To fill this gap, we perform a first-of-its-kind holistic security and privacy risk assessment of ML models, vis-\`a-vis four representative inference attacks.

\mypara{Threat Model Taxonomy}
Our work starts with a systematical categorization of the knowledge that an adversary might have to launch the attacks, along two dimensions: 1) access to a target model (white-box or black-box), 2) availability of an auxiliary dataset (partial training dataset, shadow dataset, or no dataset).
We consider four types of state-of-the-art inference attacks and describe under which threat models they can be applied.
This provides us with a comprehensive spectrum of the inference attack surface for ML models.

\mypara{Experimental Evaluation}
We perform an extensive measurement study of the attacks, jointly, over five different ML model architectures (AlexNet~\cite{KSH12}, ResNet18~\cite{HZRS16}, VGG19~\cite{SZ15}, Xception~\cite{C17}, and SimpleCNN) and four image datasets (CelebA~\cite{LLWT15}, Fashion-MNIST (FMNIST)~\cite{XRV17}, STL10~\cite{CNL11}, and UTKFace~\cite{ZSQ17}).
Our analysis aims to answer three research questions: 
1) What is the impact of dataset complexity on different attacks?
2) What is the impact of overfitting on different attacks?
3) What is the relationship among different attacks?

\mypara{Main Findings} 
Our analysis shows that the complexity of the target model's training dataset plays a major role in the accuracy of membership inference, model inversion, and model stealing.
In particular, membership inference is much more effective on complex datasets, while the other two exhibit the opposite trend.
For instance, for membership inference (with black-box access to the target model and a shadow dataset) against ResNet18, there is a 68.4\% increase when going from a simple dataset (FMNIST) to a complex one (STL10).\footnote{We refer to both sample complexity and class diversity (see \autoref{section:effectdataset}).} 
On the other hand, model stealing achieves 0.524 agreement (the standard metric for this attack) on ResNet18 trained on STL10 but a much higher 0.927 on FMNIST.
This stems from ML models being more prone to overfitting on complex datasets, which leads to better membership inference, whereas when an ML model is trained on a complex dataset, it is harder for an adversary to obtain a dataset with similar complexity (by querying the target model) to train their stolen model.

We also find that the performance of membership inference and model stealing are negatively correlated ($r=-0.821$), i.e., a target model with higher membership risks is less vulnerable to model stealing.
This is due to the opposite effect of overfitting on these two attacks.
Moreover, access to a partial training dataset does not significantly improve attack performance for membership inference, attribute inference, and model stealing.
E.g., the accuracy for attribute inference (on ResNet18/CelebA) is 0.719 with a partial training dataset and 0.726 with a shadow dataset.

\mypara{Defenses} 
We then evaluate two defense mechanisms, i.e., DP-SGD~\cite{ACGMMTZ16} and Knowledge Distillation (KD)~\cite{HVD15}, against all the inference attacks.
Empirical results show that DP-SGD can mitigate membership inference attacks in general without damaging target models' utility significantly.
Meanwhile, KD also reduces membership inference risks, but generally, to a lesser extent compared to DP-SGD.
However, neither of them is effective against other inference attacks.
This highlights the lack of a general, effective defense mechanism, and our work sheds light as to what extent/why.

\mypara{\systemName}
To support the comprehensive evaluation of a wide range of inference attacks and defenses (current and future), we introduce a framework called \systemName.
This can be used by multiple entities and for multiple purposes.
For instance, model owners can use it to seamlessly and meaningfully assess potential security and privacy risks before deploying their model.
Also, as we make source code publicly available, researchers will be able to re-use \systemName to benchmark new inference attacks and defense mechanisms.
\systemName follows a modular design, which easily supports the integration of additional inference attacks and defenses, as well as plugging in datasets, models, etc.

\section{Threat Modeling}
\label{section:threat}

In this work, we focus on image classification, one of the most popular ML applications.
In general, the goal of an ML classifier is to map a data sample to a label/class.
The input to an ML model is a data sample, and the output is a vector of probabilities, or posteriors, with each element representing the likelihood of the sample belonging to a class.

We categorize the threat models for all the inference attacks considered in this paper along two dimensions, i.e., 1) \textit{access to the target model} and 2) \textit{auxiliary dataset}.
In total, we consider five different scenarios.

\mypara{Access to the Target Model}
We consider two access settings: \textit{white-box} and \textit{black-box}. 
The former, denoted with \WM, means that an adversary has full information about the target model, including its parameters and architecture. 
In black-box attacks, denoted with \BM, the adversary can only access the target model in an API-like manner, e.g., they can query the target model and get the model's output.
However, most of the existing black-box literature~\cite{SSSS17,GWYGB18,XWLBGL21} also assumes that the adversary knows the target model's architecture which they use to build shadow models (see \autoref{section:attack}).

Overall, the white-box model captures scenarios where the target model's parameters are leaked, e.g., following a data breach or through reverse engineering, e.g., from pre-trained models deployed to mobile devices~\cite{HVD15}.
The black-box model encapsulates API access akin to features provided by Machine Learning as a Service (MLaaS) platforms.

\mypara{Auxiliary Dataset}
The adversary needs an auxiliary dataset in order to train their attack model.
We consider three scenarios, in decreasing order of adversarial ``strength'': 1) \textit{partial training dataset} (\PD), 2) \textit{shadow dataset} (\SD), and 3) \textit{no dataset} (\ND). 
In the first scenario, the adversary obtains parts of the actual training data from the target model (e.g., it is public knowledge), while in the last one, they have no information at all.
In between is the \SD setting, where the adversary gets a ``shadow'' dataset from the same distribution as the target model's training data (see Section V-C in~\cite{SSSS17} for a discussion on how to generate such data, using, e.g., through model-based or statistics-based synthesis, or noisy real data).

\mypara{Considered Settings} 
Overall, the two different types of model access and the three types of auxiliary dataset availability lead to six threat models.
In the rest of the paper, we consider five of them: $\tuple{\BM, \PD}$, $\tuple{\BM, \SD}$, $\tuple{\WM, \PD}$, $\tuple{\WM, \SD}$, and $\tuple{\WM, \ND}$.
We do not experiment with black-box access and no auxiliary dataset, as this is unlikely to yield successful attacks. 

\section{Inference Attacks}
\label{section:attack}

In this section, we present the four inference attacks measured in this paper.
Specifically, we consider membership inference (\meminf), model inversion (\modelinv), attribute inference (\attrinf), and model stealing (\modelsteal).
The first three are designed to infer information about a target ML model's training data, while the last one aims to steal the target model's parameters.

\begin{table}[!t]
\centering
\customTableFont
\setlength{\tabcolsep}{10 pt}
\begin{tabular}{@{}l@{}l | c c@{}}
\toprule
\multicolumn{2}{c|}{\bf Auxiliary} & \multicolumn{2}{c}{\bf Model Access}\\
\multicolumn{2}{c|}{\bf Dataset} & Black-Box (\BM) & White-Box (\WM) \\
\midrule
Partial & (\PD) & \meminf, \modelsteal & \meminf, \attrinf \\
Shadow~ &(\SD) & \meminf, \modelsteal & \meminf, \attrinf, \modelinv \\
No & (\ND) & - & \modelinv\\
\bottomrule
\end{tabular}
\caption{Different attacks under different threat models.}
\label{table:taxonomy}
\end{table}

Different attacks can be applied to different threat models; see \autoref{table:taxonomy}.
For each attack and each threat model, we concentrate on one representative state-of-the-art method.

\subsection{Membership Inference}

Membership Inference (\meminf)~\cite{SSSS17} against ML models involves an adversary aiming to determine whether or not a target data sample was used to train a target ML model.
More formally, given a target data sample $\xtarget$, (the access to) a target model $\model{}{}$, and an auxiliary dataset $\dset{\aux}{}$, a membership inference attack can be defined as: \vspace{-0.2cm}
\[
\meminf:\xtarget,\model{}{},\dset{\aux}{}\rightarrow \{\textit{member}, \textit{non-member}\} \vspace{-0.2cm}
\]
where $\model{}{}\in\{\BM,\WM\}$ and $\dset{\aux}{}\in\{\PD,\SD\}$.

Membership inference has been extensively studied in literature~\cite{SSSS17,NSH18,SZHBFB19,JSBZG19,SDSOJ19,LZ21,CYZF20,LF20,CZWBHZ21}.
Inferring membership of a target sample prompts severe privacy threats; for instance, if an ML model for drug dose prediction is trained using data from patients with a certain disease, then inclusion in the training set inherently leaks the individuals' health status.
Overall, \meminf is also often a signal that a target model is ``leaky'' and can be a gateway to additional attacks~\cite{C20}.

In the following, we illustrate how to implement membership inference (\meminf) under different threat models.

\mypara{Black-Box/Shadow \atktu{\meminf}{\BM}{\SD}~\cite{SZHBFB19}}
We start with the most common and difficult setting for the attack~\cite{SSSS17,SZHBFB19}, whereby the adversary has black-box access (\BM) to the target model and a shadow auxiliary dataset (\SD).

The adversary first splits the shadow dataset into two parts and uses one to train a shadow model on the same task.
Next, the adversary uses the entire shadow dataset to query the shadow model.
For each querying sample, the shadow model returns its posteriors and the predicted label: if the sample is part of the shadow model's training set, the adversary labels it as a member and as a non-member otherwise.
With this labeled dataset, the adversary trains an attack model, which is a binary membership classifier.
Finally, to determine whether a data sample is a member of the target model's training dataset, the sample is fed to the target model, and the posteriors and the predicted label (transformed to a binary indicator on whether the prediction is correct) are fed to the attack model. 

\mypara{Black-Box/Partial \atktu{\meminf}{\BM}{\PD}~\cite{SZHBFB19}}
If the adversary has black-box access to the target model and a partial training dataset, the attack method is very similar to that for \atktu{\meminf}{\BM}{\SD}.
However, the adversary does not need to train a shadow model; rather, they use the partial training dataset as the ground truth for membership and directly train their attack model.

\mypara{White-Box/Shadow \atktu{\meminf}{\WM}{\SD}~\cite{NSH19}} 
Nasr et al.~\cite{NSH19} introduce an attack in the white-box setting with either a shadow or a partial training dataset as the auxiliary dataset.\footnote{The attack by Nasr et al.~\cite{NSH19} was originally designed for the partial training dataset setting, but it can be adapted to the shadow dataset setting.} 
In the former, similar to \atktu{\meminf}{\BM}{\SD}, the adversary uses \SD to train a shadow model to mimic the behavior of the target model and to generate data to train their attack model.
As the adversary has white-box access to the target model, they can also exploit the target sample's gradients with respect to the model parameters, embeddings from different intermediate layers, classification loss, and prediction posteriors (and label). 

\mypara{White-Box/Partial \atktu{\meminf}{\WM}{\PD}~\cite{NSH19}}
The attack methodology here is almost identical to the black-box counterpart.
The only difference is that the adversary can use the same set of features as the attack model for \atktu{\meminf}{\WM}{\SD}.

\subsection{Model Inversion}

Model inversion attacks (\modelinv)~\cite{FJR15} aim to reconstruct data samples from a target ML model, i.e., they allow an adversary to directly learn information about the training dataset.

For instance, in a facial recognition system, a \modelinv adversary tries to learn the facial data of a victim whose data is used to train the model.
Model inversion requires the adversary to have white-box access to the target model; this is due to the fact that the attack needs to perform back-propagation over the target model's parameters (detailed below).

Formally, we define model inversion as: \vspace{-0.2cm}
\[
\modelinv:\WM,\dset{\aux}{}\rightarrow \{\textit{training samples}\} \vspace{-0.2cm}
\]
where $\dset{\aux}{}\in\{\ND,\SD\}$.

We consider two types of model inversion attacks: the one proposed by Fredrikson et al.~\cite{FJR15}, which aims to reconstruct a representative sample for each class of the target model, and that by Zhang et al.~\cite{ZJPWLS20}, which aims to synthesize the training dataset.
These two attacks follow different threat models, which we discuss next.

\mypara{White-Box/No Auxiliary \atktu{\modelinv}{\WM}{\ND}~\cite{FJR15}}
The method by Fredrikson et al.~\cite{FJR15} assumes the adversary has white-box access to the target model\footnote{Fredrikson et al.~\cite{FJR15} also introduce a model inversion attack where the adversary only has black-box access to the target model; however, its performance is not as good and therefore we do not consider it.} and does not need any auxiliary dataset.
For each class of the target model, the adversary first creates a noise sample, feeds this sample to the model, and gets the posteriors.
The adversary then uses back-propagation over the target model's parameters to optimize the input sample so that the corresponding posterior of the class can exceed a pre-set threshold.
Once the threshold is reached, the optimized sample is the representative sample of that class, i.e., the attack output.

\mypara{White-Box/Shadow \atktu{\modelinv}{\WM}{\SD}~\cite{ZJPWLS20}}
The attack by Zhang et al.~\cite{ZJPWLS20} uses a shadow dataset to enhance the quality of the reconstructed samples by training a generative adversarial network (GAN)~\cite{GPMXWOCB14}.
First, the adversary trains a GAN with a shadow dataset.
Next, the adversary optimizes the inputs to the GAN's generator, i.e., the noise, to find those GAN-generated samples that can achieve high posteriors on the target model.
These samples are the attack's final outputs.
In other words, this attack performs optimization on the inputs to the GAN instead of the samples to the target model directly~\cite{FJR15}.
Since the GAN is capable of generating high-quality samples, the attack's final outputs will be more realistic.
Moreover, due to the fact that GAN can generate multiple samples, this attack is able to generate multiple samples for each class of the target model as well.

\subsection{Attribute Inference}

An ML model may learn extra information about the training data that is not related to its original task;
e.g., a model predicting age from profile photos can also learn to predict race~\cite{MSCS19,SS20}.
Attribute inference (\attrinf) aims to exploit such unintended information leakage.

State-of-the-art attacks usually rely on the embeddings of a target sample ($\xtarget$) obtained from the target model to predict the sample's target attributes.
Thus, the adversary is assumed to have white-box access to the target model.
Formally, attribute inference is defined as: \vspace{-0.15cm}
\[
\attrinf:\xtarget,\WM,\dset{\aux}{}\rightarrow \{\textit{target attributes}\} \vspace{-0.15cm}
\]
where $\dset{\aux}{}\in\{\PD,\SD\}$ can either be a partial training dataset or a shadow dataset.

\mypara{White-Box/Shadow and Partial~\cite{MSCS19,SS20}}
Both attacks, i.e., \atktu{\attrinf}{\WM}{\SD}~\cite{MSCS19} and \atktu{\attrinf}{\WM}{\PD}~\cite{SS20}, follow a similar attack methodology.
The only difference lies in the dataset used to train the attack model.
In both cases, the adversary is assumed to know the target attributes of the auxiliary dataset.
Then, they use the embeddings, and the target attributes to train a classifier to mount the attack.

\subsection{Model Stealing}

The goal of model stealing attacks (\modelsteal)~\cite{TZJRR16,OSF19}, aka model extraction, is to extract the parameters from a target model.
Ideally, an adversary will be able to obtain a model (the ``stolen'' model) with very similar performance as the target model.
More formally: \vspace{-0.15cm}
\[
\modelsteal:\BM,\dset{\aux}{}\rightarrow \model{}{\mathsf{C}} \vspace{-0.15cm}
\]
where \model{}{\mathsf{C}} is the stolen model and $\dset{\aux}{}\in\{\PD,\SD\}$.

Model stealing prompts severe security risks.
For instance, as it is often difficult to train an advanced ML model (e.g., due to the lack of data or computing resources), stealing a trained model inherently constitutes intellectual property theft.
Also, as many other attacks, such as adversarial examples~\cite{PMGJCS17}, require white-box access to the target ML model, model stealing can be a stepping stone to perform these attacks.

\mypara{Black-Box/Partial and Shadow~\cite{TZJRR16}}
In this paper, we concentrate on the attacks proposed by Tram{\`e}r et al.~\cite{TZJRR16}, for \atktu{\modelsteal}{\BM}{\PD} and \atktu{\modelsteal}{\BM}{\SD}.
The adversary is assumed to have knowledge of the target model's architecture, and both attacks follow a similar methodology.
Specifically, the adversary uses data samples from their auxiliary dataset (\PD or \SD) to query the target model and get the corresponding posteriors.
Then, they use them to train the stolen model, with the posteriors as the ground truth.

\section{\systemName}
\label{section:system}

In this section, we introduce \systemName, a modular framework geared to evaluate the four inference attacks, as well as the two defenses (see \autoref{section:defense}), considered in this paper.

Prior work has proposed software tools to evaluate attacks against ML models, such as adversarial examples~\cite{PFCGFKXSBRMBHZJLSGUGDBHRLM18,LJZWWLW19}, backdoor attacks~\cite{PZGXJCW20}, and membership inference~\cite{MS20}.
To the best of our knowledge, \systemName is the first framework that jointly considers different types of inference attacks.

\begin{figure}[!t]
\centering
\includegraphics[width=\columnwidth]{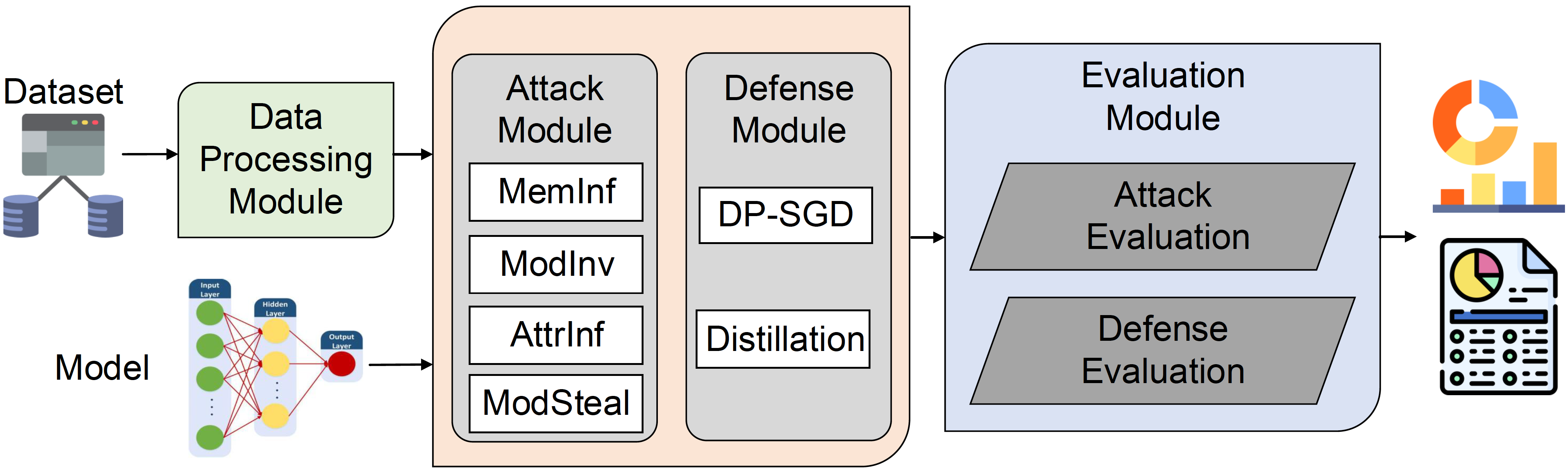}
\caption{Overview of \systemName's modules.}
\label{figure:system}
\vspace{-0.15cm}
\end{figure} 

\mypara{Modules}
In \autoref{figure:system}, we report the four different modules of \systemName:
\begin{enumerate}
\itemsep-0.28em 
\item \textbf{Data Processing.} 
This module processes the datasets to mount different attacks.
It also involves data pre-processing methods, e.g., normalization.
\item \textbf{Attack.} This module performs the actual inference attacks.
At the moment, it supports ten different attacks belonging to four different attack types (see \autoref{section:attack}).
\item \textbf{Defense.} We currently support two representative mitigation techniques for inference attacks against ML models, as discussed later in \autoref{section:defense}.
\item \textbf{Evaluation.} This module is used to evaluate the performance of attacks and defenses. 
\end{enumerate}
The modular design of \systemName allows to easily integrate additional attacks and defense mechanisms, as well as plugging in any dataset or model.

\mypara{Using \systemName}
A user needs to input their target model and its training dataset to use \systemName.
This is to achieve a full-fledged privacy risk assessment.
We envision \systemName to be used for the following purposes:
\begin{itemize}
\itemsep-0.28em 
\item As it supports a systematic taxonomy of different threat models for inference attacks, \systemName enables model owners to obtain an overview of the threats their model may face when deployed in the real world.
\item \systemName provides a holistic assessment of different attacks, as well as the effectiveness of possible defenses.
To our best knowledge, this is the first tool to provide such a comprehensive analysis of inference attacks.
\item Researchers can re-use \systemName as a benchmark tool to experiment with new inference attacks and defenses in the future.
\systemName's data processing and evaluation modules can be seamlessly re-used by other attacks or defenses.
Moreover, the maturity of the topic, as demonstrated by the state of the art~\cite{SSSS17,SZHBFB19,FJR15,TZJRR16,ZJPWLS20,SS19,KTPPI20}, suggests that new attacks against ML models are very likely to fall into one of the threat models summarized in our taxonomy (\autoref{section:threat}).
This means that new attacks/defenses can be easily implemented within \systemName's attack and defense modules.
\item Since \systemName follows a modular design, the communication between different modules is implemented using an API-based approach. 
To include a new attack/defense, one only needs to specify the attack/defense models' architecture in the attack/defense module, which can be easily done with the support of the current deep learning libraries.
To further extend \systemName into different domains like text or audio, users can re-implement the processing function and attack methods in the corresponding modules, and reuse other modules directly.
\end{itemize}

\section{Experimental Settings}
\label{section:experiments}

\subsection{Experimental Protocol}

We first select four benchmark datasets (see \autoref{section:dataset}) and five state-of-the-art ML models (see \autoref{section:target_model}) to train a total of 20 target models.
These are used to evaluate different attacks (see \autoref{section:attack}) and defenses (see \autoref{section:defense}).
For each dataset, we partition it into four parts (see \autoref{section:dataset}), including target training dataset, target testing dataset, shadow training dataset, and shadow testing dataset, to comply with the different threat models discussed in \autoref{section:threat}.

We then submit each target model and the corresponding dataset partition to \systemName, running the attacks (see \autoref{section:attackmodels}) and applying the defenses (see \autoref{section:defense}). 
Finally, we use the evaluation module of \systemName to summarize the results and perform a comprehensive analysis to answer the research questions listed in \autoref{section:question}.

\subsection{Datasets}
\label{section:dataset}

For the sake of this paper, we experiment with four datasets: 

\begin{itemize}
\itemsep-0.28em 
\item \textbf{CelebA~\cite{LLWT15}} contains 202,599 face images, each is associated with 40 binary attributes. 
We select and combine 3 attributes out of 40, including \textit{HeavyMakeup}, \textit{MouthSlightlyOpen}, and \textit{Smiling} to form our target models' classes/labels, leading to 8-class classification.
\item \textbf{FMNIST (Fashion-MNIST)~\cite{XRV17}} is also an image dataset containing 70,000 gray-scale images equally distributed among 10 different classes, including T-shirt, trouser, pullover, dress, coat, sandal, shirt, sneaker, bag, and ankle boot.
\item \textbf{STL10~\cite{CNL11}} is a 10-class image dataset, each contains 1,300 images.
The classes include airplane, bird, car, cat, deer, dog, horse, monkey, ship, and truck.
\item \textbf{UTKFace~\cite{ZSQ17}} has 23,000 face images associated with age, gender, and race. 
We consider the images from the largest four races (White, Black, Asian, and Indian) in the dataset and use race as the label for the corresponding target models.
This leaves us with 22,012 images.\vspace{-0.15cm}
\end{itemize}
Note that all the samples in the datasets are re-sized to 32$\times$32 pixels.
This is common practice in ML and ensures that the comparison among different datasets is fair.
We randomly split each dataset into four equal disjoint parts:

\begin{enumerate}
\itemsep-0.28em 
\item \textbf{Target Training Dataset} is used to train all the target models and to evaluate the performance of all membership inference attacks and model inversion attacks.
For attacks that require a partial training dataset (\PD), i.e., \atktu{\meminf}{\BM}{\PD}, \atktu{\meminf}{\WM}{\PD}, \atktu{\attrinf}{\WM}{\SD}, \atktu{\attrinf}{\WM}{\PD}, and \atktu{\modelsteal}{\BM}{\PD}, we randomly select 70\% samples from the target training dataset.
\item \textbf{Target Testing Dataset} is used to evaluate the performance of the target model.
It is also used to evaluate the performance of all membership inference, attribute inference, and model stealing attacks.
\item \textbf{Shadow Training Dataset} is used to train all the attack models that require a shadow auxiliary dataset.
\item \textbf{Shadow Testing Dataset} is used to train two membership inference attack models, i.e., \atktu{\meminf}{\BM}{\SD} and \atktu{\meminf}{\WM}{\SD}, that require a shadow dataset as the auxiliary dataset.
\end{enumerate}

In a nutshell, all the datasets we choose in this paper are benchmark datasets for evaluating inference attacks against ML models ~\cite{SS19,LJQG21,HZ21,JLG22,KD21}.
These datasets have different numbers of classes and cover a variety of objects, e.g., human faces, transportation tools, and animals. 
Also, the images are ranging from gray-scale to colored in different datasets.
Note that \systemName is not bounded by certain types of datasets.
We plan to extend \systemName to other security-related datasets such as network scans, malware traces, etc.

\subsection{Target Models} 
\label{section:target_model}

We focus on five model architectures, including AlexNet~\cite{KSH12}, ResNet18~\cite{HZRS16}, VGG19~\cite{SZ15}, Xception~\cite{C17}, and SimpleCNN (containing 2 convolutional layers and 2 fully connected layers) for all the four datasets introduced above.
In total, we train 20 different target models.

For training, we set the mini-batch size to 64 and use cross-entropy as the loss function.
We use stochastic gradient descent (SGD) as the optimizer with a weight decay of 5e-4 and momentum of 0.9.
Each target model is trained for 300 epochs.
The learning rate is 1e-2 before 50 epochs, 1e-3 from 50-100 epochs, and 1e-4 until the end.
All target models' training and testing accuracy are shown in \autoref{table:model_acc}.
Note that for shadow models used in the membership inference attacks, we train them following the same process as the target models.

\subsection{Attack Models}
\label{section:attackmodels}

\mypara{Membership Inference} 
Recall that there are four different scenarios for \meminf; we establish two types of attack models: one for the black-box and the other for the white-box setting.
For black-box, our attack model has two inputs; the target sample's ranked posteriors and a binary indicator on whether the target sample being predicted correctly.
Each input is first fed into a different 2-layer MLP (Multilayer Perceptron), then the two obtained embeddings are concatenated together and fed into a 4-layer MLP.
For white-box, we have four inputs for this attack model, like the one used by Nasr et al.~\cite{NSH19}, including the target sample's ranked posteriors, classification loss, gradients of the parameters of the target model's last layer, and one-hot encoding of its true label.
Each input is fed into a different neural network, and the resulted embeddings are concatenated together as input to a 4-layer MLP.
We use ReLU as the activation function for the attack models.
The mini-batch size is set to 64, and cross-entropy is the loss function.
We use Adam as the optimizer, with learning rate of 1e-5.
The attack model is trained for 50 epochs.
We adopt {\em accuracy, F1 score, and AUC (area under the ROC curve) score} as the evaluation metrics.

\mypara{Model Inversion} 
For \atktu{\modelinv}{\WM}{\ND}, following the attack settings in~\cite{FJR15}, we set the threshold to 0.999, learning rate to 1e-2, maximum iteration to 3,000, and early stop criteria to 100.
This attack can only generate one representative sample for each class of the target model.
To evaluate the quality of the reconstructed sample, we first obtain an average sample from all samples of each target class, then calculate the mean squared error (MSE) between this average sample and the reconstructed sample.
Finally, we use the average of the MSE values for all target classes as the evaluation metric.
Note that smaller MSE within the same dataset indicates better attack performance.
However, for different datasets, different MSE can be caused by the different normalization effects.
For example, FMNIST has the highest MSE; this is due to the characteristic of FMNIST: most of the pixels are normalized to -1 or 1.

For \atktu{\modelinv}{\WM}{\SD}~\cite{ZJPWLS20}, we first use the shadow training dataset to train a DCGAN~\cite{RMC15} with the generator's noise dimension setting to 100.
For the attack, we set the learning rate to 1e-3, momentum to 0.9, loss ratio $\lambda$ to 100, iteration round to 1,500, and clip range to 1.
To evaluate the effectiveness of this attack, we use the same approach by Zhang et al.~\cite{ZJPWLS20}, i.e., we train an evaluation classifier on the identical task of the target model and use this evaluation classifier to check whether the reconstructed samples can be recognized correctly.
We use {\em accuracy and macro-F1 score} of this evaluation classifier on reconstructed samples as the performance metrics.

\mypara{Attribute Inference} 
We only use two datasets, namely, CelebA and UTKFace, to evaluate this attack as both of them have extra attributes that can be used as the target attributes.
For the former, we utilize \textit{Male}/\textit{Female} and \textit{Young}/\textit{Old} as the target attribute, resulting in a combination of four target attribute values.
For the latter, we choose \textit{Male}/\textit{Female} as the target attribute.

Our attack model is a 2-layer MLP; its input is the target sample's embeddings from the second-to-last layer of the target model. 
We use cross-entropy as the loss function and Adam as the optimizer with learning rate of 1e-3.
The attack model is trained for 50 epochs.
For the evaluation metrics, we use {\em accuracy} and {\em F1 score} ({\em macro-F1 score} for CelebA as the attack has four target attributes).

\mypara{Model Stealing}
We evaluate the model stealing attack over all the 20 target models.
For the stolen model, we use the same architecture as the target model~\cite{TZJRR16}.
Each stolen model is trained using the MSE loss and SGD as the optimizer (momentum 0.9 and learning rate 1e-2) for 50 epochs.
{\em Accuracy and agreement} are used to assess the success of the attack, where agreement represents the proportion of samples in the target testing dataset on which the target and the stolen models make the same prediction.

\begin{table}[!t]
\centering
\setlength{\tabcolsep}{4 pt}
\customTableFont
\begin{tabular}{@{}l| c c  c c@{}}
\toprule
& {\bf CelebA} & {\bf FMNIST} & {\bf STL10} & {\bf UTKFace} \\
\midrule
{\bf AlexNet} & 1.000 / 0.680 & 1.000 / 0.884 & 1.000 / 0.522 & 1.000 / 0.792\\
{\bf ResNet18} & 1.000 / 0.742 & 1.000 / 0.909 & 1.000 / 0.524 & 1.000 / 0.852\\
{\bf VGG19} & 1.000 / 0.734 & 1.000 / 0.905 & 1.000 / 0.587 & 1.000 / 0.834\\
{\bf Xception} & 1.000 / 0.735 & 1.000 / 0.916 &  1.000 / 0.574 & 1.000 / 0.846\\
{\bf SimpleCNN} & 1.000 / 0.707 & 1.000 / 0.903 & 1.000 / 0.517 & 1.000 / 0.818\\
\bottomrule
\end{tabular}
\caption{Performance of target models, namely, training/testing accuracy for each setting.}
\label{table:model_acc}
\vspace{-0.15cm}
\end{table}

\begin{figure*}[!t]
\centering
\includegraphics[width=1.9\columnwidth]{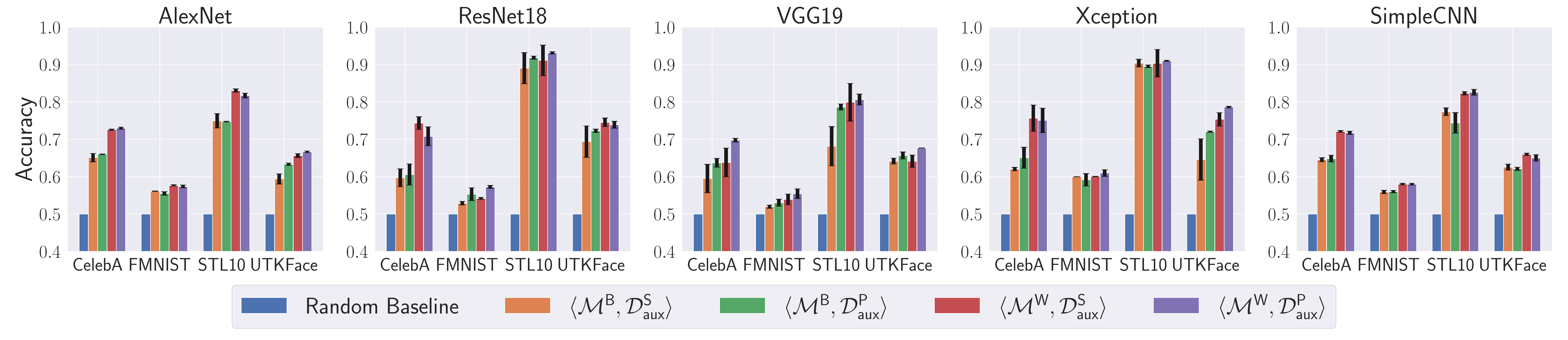}
\caption{Accuracy of membership inference attacks (\meminf) under different threat models, datasets, and target model architectures.}
\label{figure:meminf}
\end{figure*} 

\begin{figure*}[!t]
\centering
\includegraphics[width=1.9\columnwidth]{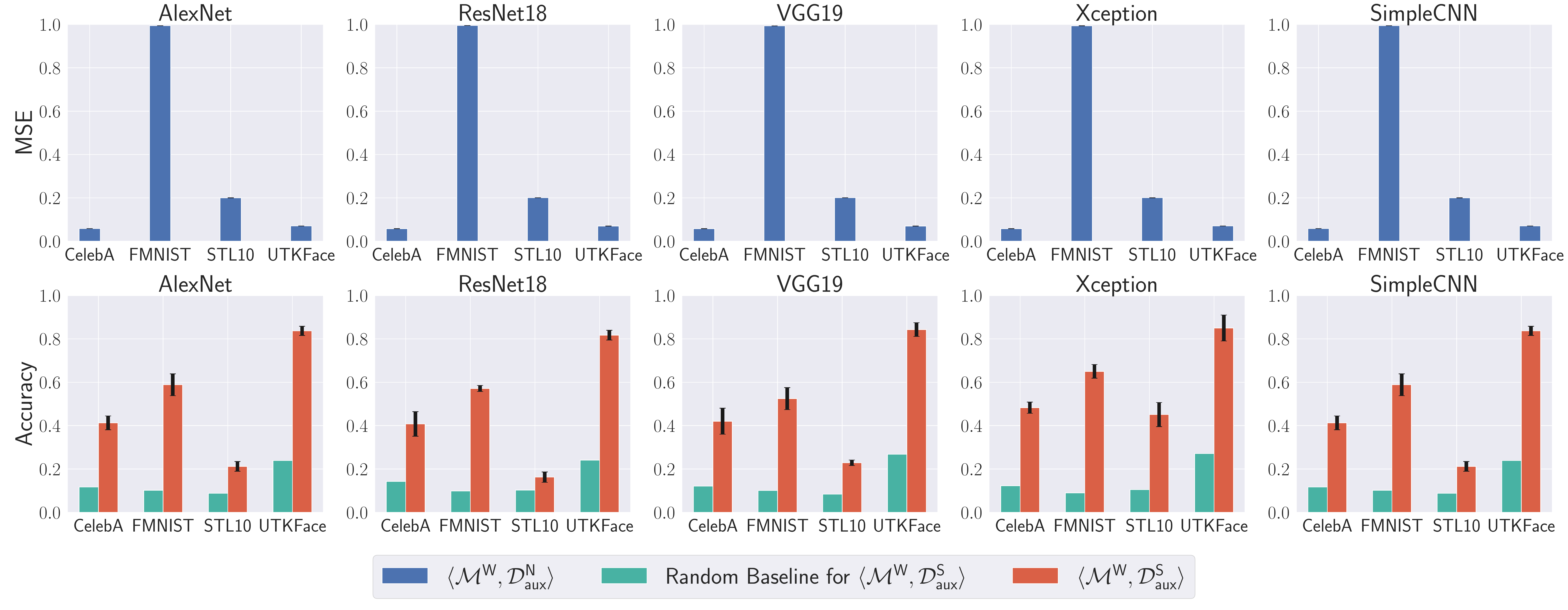}
\caption{MSE (\atktu{\modelinv}{\WM}{\ND}) and accuracy (\atktu{\modelinv}{\WM}{\SD}) of model inversion attacks (\modelinv) under different threat models, datasets, and target model architectures.}
\label{figure:modinv}
\end{figure*} 

\section{Experimental Evaluation} 
\label{section:evaluation}

In this section, we build on \systemName to provide a holistic assessment of inference attacks against ML models.
Experiments are performed on an NVIDIA DGX-A100 server with Ubuntu 18.04 operating system.
We run all the experiments 10 times, reporting mean and standard deviation values.

\subsection{Research Questions}
\label{section:question}

We start by assessing the overall performance of the four attacks.
We then analyze the impact of dataset and overfitting on the attack performance, as well as the relationship among different attacks.
Concretely, we aim to answer the following key research questions:
\begin{itemize}
\itemsep-0.28em 
\item {\em RQ1:} What is the impact of dataset complexity on different attacks?
\item {\em RQ2:} What is the impact of overfitting on different attacks?
\item {\em RQ3:} What is the relationship among different attacks?
\end{itemize}

\subsection{Attack Performance}
\label{section:attackresult}

\begin{figure*}[!t]
\centering
\includegraphics[width=1.9\columnwidth]{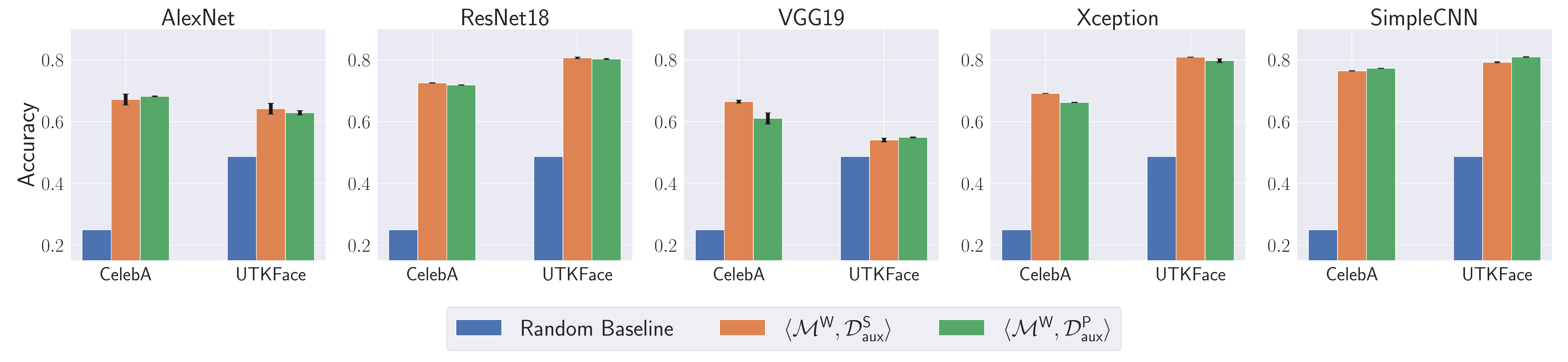}
\caption{Accuracy of attribute inference attacks (\attrinf) under different threat models, datasets, and target model architectures.}
\label{figure:attrinf}
\end{figure*}

\begin{figure*}[!t]
\centering
\includegraphics[width=1.9\columnwidth]{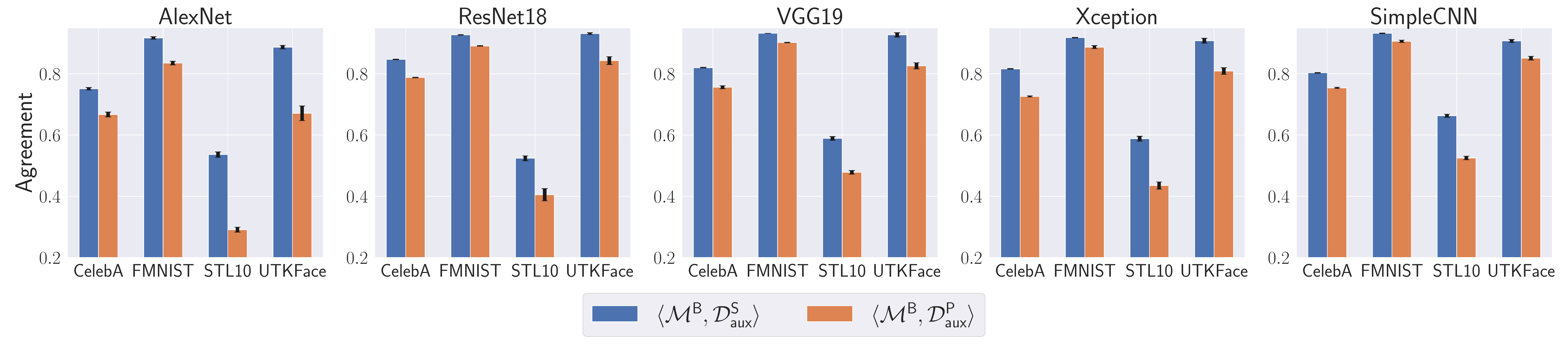}
\caption{Agreement of model stealing attacks (\modelsteal) under different threat models, datasets, and target model architectures.}
\label{figure:modsteal}
\end{figure*} 

\begin{figure*}[!t]
\centering
\includegraphics[width=1.9\columnwidth]{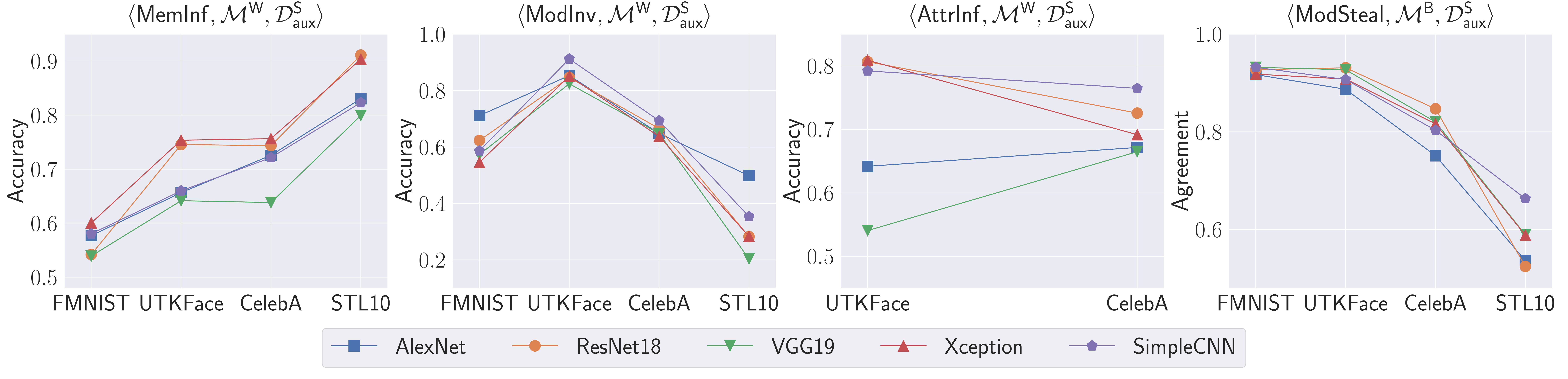}
\caption{The relation between dataset complexity and attack performance.
For \meminf, \modelinv, and \attrinf. we use accuracy, for \modelsteal, agreement.}
\label{figure:atkvsdata}
\end{figure*} 

\begin{figure*}[!t]
\centering
\includegraphics[width=1.9\columnwidth]{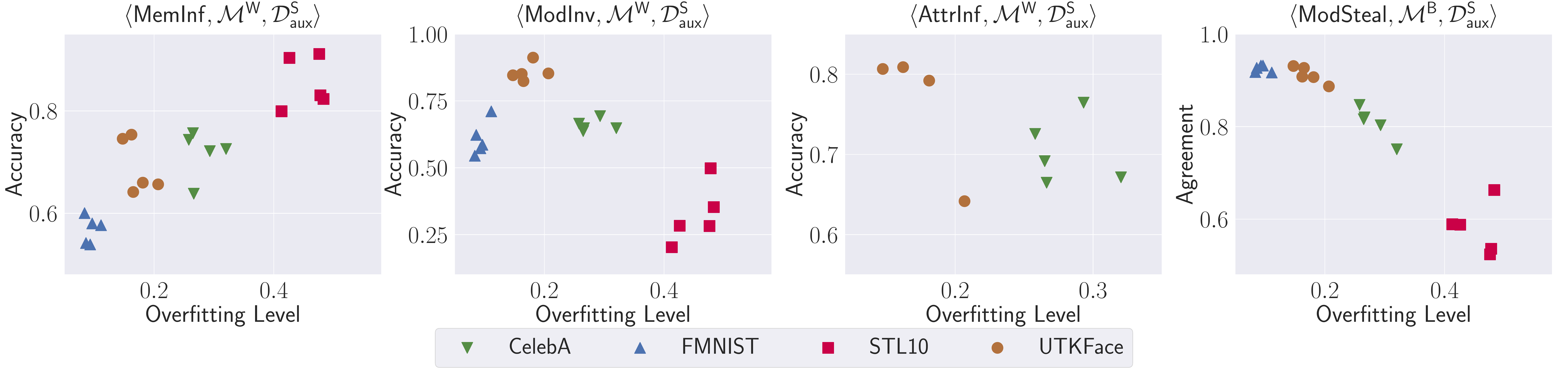}
\caption{The relation between overfitting level (on target models) and attack performance. 
For \meminf, \modelinv, and \attrinf, we use accuracy, for \modelsteal, agreement.}
\label{figure:overfiting}
\end{figure*}

\mypara{Membership Inference}
In~\autoref{figure:meminf}, we report the accuracy of \meminf.
We observe that the attack achieves high accuracy on CelebA, STL10, and UTKFace.
For instance, the attack accuracy of \atktu{\meminf}{\WM}{\SD} on ResNet18 trained on the STL10 dataset is 0.911.
On the other hand, the performance on FMNIST is not strong, as models trained on FMNIST generalize well on non-member data samples---in other words, there is less overfitting~\cite{SSSS17} (see \autoref{section:effectoverfitting}).
We also report the F1 score and AUC score in \autoref{section:appendix} (\autoref{figure:meminf_f1} and \autoref{figure:meminf_auc}), and the corresponding ROC curves are depicted in \autoref{section:appendix} (\autoref{figure:roc_celeba}, \autoref{figure:roc_FMNIST}, \autoref{figure:roc_stl10}, and \autoref{figure:roc_UTKFace}).

An adversary with white-box access to the target model generally achieves better performance than the one with black-box access.
For instance, the accuracy of \atktu{\meminf}{\WM}{\SD} is higher than that of \atktu{\meminf}{\BM}{\SD}, except for Xception on STL10 and FMNIST and VGG19 on UTKFace (see \autoref{figure:meminf}).
A similar observation can be drawn from \atktu{\meminf}{\WM}{\PD} and \atktu{\meminf}{\BM}{\PD}; this is expected as the adversary can exploit more information in the white-box setting.
In particular, we find that the classification loss possesses the strongest signal among others for the attack~\cite{NSH19}.
Meanwhile, partial training dataset also leads to better membership inference performance than the shadow dataset; however, the effect is less pronounced. 

\begin{figure*}[!t]
\centering
\includegraphics[width=1.9\columnwidth]{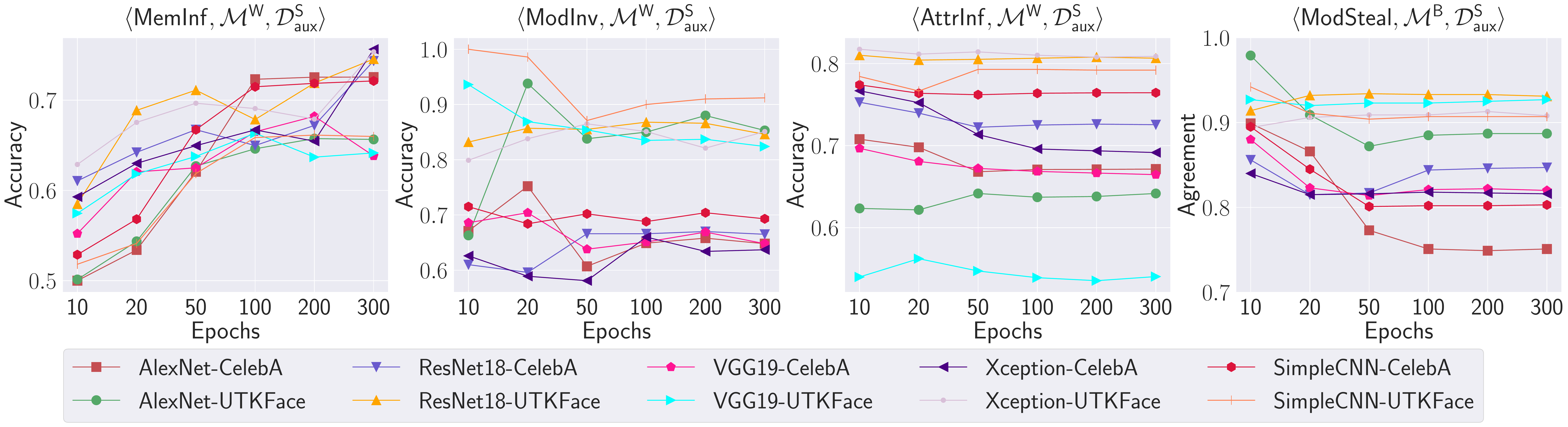}
\caption{The relation between the number of epochs and attack performance.
For \meminf, \modelinv, and \attrinf, we use accuracy, for \modelsteal, agreement.}
\label{figure:epochs}
\end{figure*}

\mypara{Model Inversion}
Next, we measure the performance of model inversion (see \autoref{figure:modinv}).
As discussed in~\autoref{section:attackmodels}, we use different metrics to evaluate these two attacks, i.e., MSE for \atktu{\modelinv}{\WM}{\ND} and accuracy for \atktu{\modelinv}{\WM}{\SD}, due to their different design.
Thus, we cannot directly compare them.
Rather, we evaluate their attack performance qualitatively (see~\autoref{figure:ccs15reverse} in \autoref{section:appendix} for two examples) and discover that the images generated by \atktu{\modelinv}{\WM}{\SD} are more realistic than those by \atktu{\modelinv}{\WM}{\ND}.
This is due to the capability of GAN for generating high-quality samples. 
For \atktu{\modelinv}{\WM}{\SD}, we also use macro-F1 as the metric; see~\autoref{figure:modinv_f1} in \autoref{section:appendix}.

\mypara{Attribute Inference}
The accuracy of the attribute inference attacks is shown in~\autoref{figure:attrinf}.
We also report macro-F1 score for CelebA and F1 score for UTKFace in~\autoref{figure:attrinf_f1} (see \autoref{section:appendix}).
The corresponding ROC curves are reported in \autoref{figure:roc_utkface_attrinf} (see \autoref{section:appendix}).
In general, the attacks work quite well.
For instance, both \atktu{\attrinf}{\WM}{\SD} and \atktu{\attrinf}{\WM}{\PD} reach around 0.800 accuracy for ResNet18 trained on UTKFace.
The F1 scores with respect to these two models are both about 0.779.
We also notice that using a partial training dataset does not provide the adversary with many advantages compared to using a shadow dataset.
In some cases, partial training dataset even yields worse performance, as in the case of VGG19 trained on CelebA.

\mypara{Model Stealing}
We report the agreement (\autoref{figure:modsteal}) and accuracy (\autoref{figure:modsteal_acc} in \autoref{section:appendix}) to evaluate model stealing attacks.
Overall, \modelsteal has strong performance.
For instance, \atktu{\modelsteal}{\BM}{\SD} for ResNet18 trained on FMNIST achieves an agreement of 0.927.
Similar to attribute inference, we observe that using a partial training dataset as the auxiliary dataset has a lower performance than the shadow dataset for model stealing.
One reason might be that using a partial training dataset querying the target model results in more confident posteriors (low entropy), which contain less information for the adversary to exploit.

\subsection{The Role of the Datasets}
\label{section:effectdataset}

To answer the first research question, we plot the relationship between dataset complexity and attack performance in~\autoref{figure:atkvsdata}.
(The x-axis represents the datasets and the y-axis shows the attack performance, and each node corresponds to one attack against one target model).
Due to space limitations, we only show one plot for one threat model for each attack.

\mypara{Dataset Complexity}
As mentioned before, all the samples in the four datasets are re-sized to 32$\times$32 pixels.
FMNIST is the simplest dataset as it only contains gray-scale images, followed by UTKFace, which consists of (full-color) human faces, and CelebA, which has 10 times more images than UTKFace.
The most complex dataset is STL10, as it contains images of 10 diverse classes, ranging from cat to ship.

\mypara{Results}
Overall, the complexity of the dataset does have a significant effect on \meminf and \modelsteal.
More precisely, more complex datasets lead to better membership inference but worse model stealing performance.
Ostensibly, this is due to the fact that a complex dataset is harder for a model to generalize on, and thus more prone to overfitting, which results in better membership inference attack~\cite{SSSS17}, whereas when a model is trained on a complex dataset, it is harder for an adversary to obtain a dataset with similar complexity (by querying the target model) to train their stolen model.

We also observe that model inversion is less effective on STL10 than on UTKFace and CelebA, whereas there is no strong influence of dataset complexity on attribute inference; this might be due to the different target classes of our attacks on these two datasets (see \autoref{section:attackmodels}).
Note that we also investigate the complexity of the target model structure on the attack performance but do not observe any clear relation.

\subsection{The Effect of Overfitting} 
\label{section:effectoverfitting}

To answer the second research question, we analyze the effect of target models' overfitting on inference attacks' performance.
Concretely, we adopt two metrics to quantify overfitting in each target model: 1) the difference between the training accuracy and the testing accuracy of the target model, referred to as the \textit{overfitting level}, and 2) the number of epochs used to train the target model~\cite{SZHBFB19}.

\mypara{Overfitting Level}
The relation between overfitting level and attack performance is shown in \autoref{figure:overfiting}.
First, we observe that different datasets have different overfitting levels, and this correlates well with the dataset complexity (see \autoref{section:effectdataset}).
Specifically, the largest (smallest) overfitting level happens on the most (least) complex dataset in our experiments, i.e., STL10 (FMNIST).

\begin{figure*}[!t]
\centering
\includegraphics[width=2.1\columnwidth]{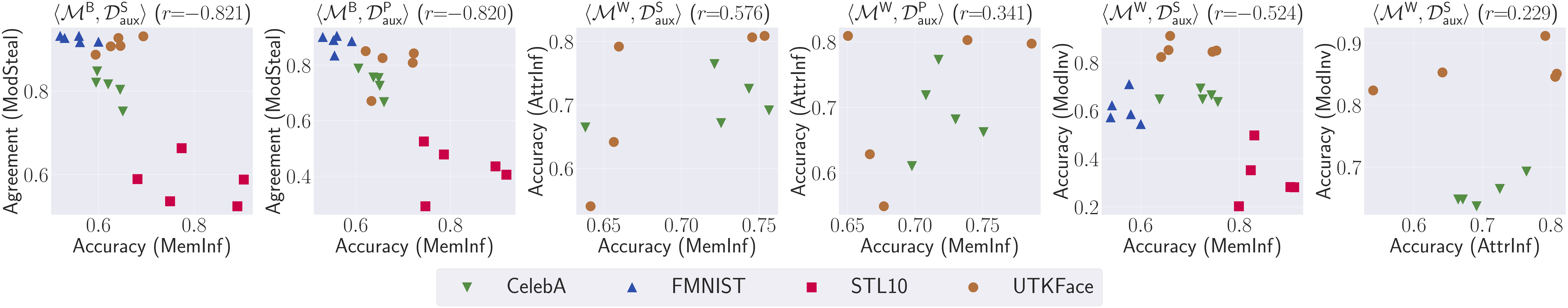}
\caption{The relation between different attacks under the same threat model.
For \meminf, \modelinv, and \attrinf, we use accuracy, for \modelsteal, agreement.}
\label{figure:correlation}
\end{figure*}

Overall, overfitting does have a significant impact on \meminf (\atktu{\meminf}{\WM}{\SD}).
That is, a higher overfitting level leads to better membership inference.
This is in line with previous analysis~\cite{SSSS17,SZHBFB19}, and is expected, as an overfitted model provides more confident predictions on its member samples (reflected on the posteriors) than on non-member samples, which can be exploited by the attack model to effectively differentiate them.

Meanwhile, model stealing displays a completely opposite trend, i.e., it is more difficult to steal a highly overfitted model.
This can be explained by the fact that an overfitted model memorizes its training dataset to a large extent, and an adversary usually does not have the ability to get the exact training dataset; thus, the stolen model is likely to be dissimilar to the target model.
Also, model inversion tends to have better performance on less overfitted models, except for FMNIST.
We believe this is due to the quality of the GAN employed in the attack.
For attribute inference, we do not observe a clear relationship between attack performance and overfitting level.

\mypara{Number of Epochs}
The relation between the number of epochs and attack performance (on UTKFace and CelebA) is shown in~\autoref{figure:epochs}.
First, we find that all attacks' performance becomes steady after 100 epochs; this is reasonable since 100 epochs are usually enough to train good target models, and further training does not cause an obvious effect on overfitting.
Second, the performance of membership inference increases from 10 epochs until 100 epochs, while model stealing shows the opposite trend.
This observation echoes our previous argument that a highly overfitted model is easier to be attacked by membership inference but harder to be stolen.
For model inversion and attribute inference, the attack performance only has slight fluctuations with a different number of epochs.

\subsection{Relation Among Different Attacks}
\label{section:attackcorrelation}

Next, we analyze the relationship between different attacks under the same threat model, which corresponds to our third research question.
In total, we consider all the six pairs of attacks (as depicted in \autoref{table:taxonomy}) including
\meminf and \modelsteal under \tuple{\BM, \SD}, \meminf and \modelsteal under \tuple{\BM, \PD}, \meminf and \attrinf under \tuple{\WM, \SD}, \meminf and \attrinf under \tuple{\WM, \PD}, \meminf and \modelinv under \tuple{\WM, \SD}, and \attrinf and \modelinv under \tuple{\WM, \SD}.

\autoref{figure:correlation} shows that there is a strong negative correlation between membership inference and model stealing (\tuple{\BM, \SD}) with respect to their accuracy ($r=-0.821$).
Specifically, worse membership inference corresponds to better model stealing.
This follows from the discussion around overfitting (\autoref{section:effectoverfitting}).
The correlation with respect to other evaluation metrics is reported in \autoref{figure:correlation_auc} in \autoref{section:appendix}.
We also observe a strong negative correlation between membership inference and model inversion, except for model inversion performing worse on FMNIST than on CelebA and UTKFace.
We speculate this is due to the capability of the DCGAN used in the model inversion attack.
FMNIST contains gray-scale images, while CelebA and UTKFace are both face datasets. 
DCGAN, in general, is more effective in generating human faces~\cite{RMC15}, which results in model inversion's better performance on CelebA and UTKFace than on FMNIST.
In the future, we plan to extend the model inversion attack by using more advanced GANs, such as StyleGAN2~\cite{KLAHLA20}.

On the other hand, there does not seem to be any clear relation between attribute inference and model inversion, as well as between membership inference and attribute inference.

\section{Defenses}
\label{section:defense}

We now evaluate two representative defense mechanisms, namely Differential Privacy (DP) and Knowledge Distillation (KD), and investigate whether or not, and how effectively, they can be used to mitigate these attacks.
To the best of our knowledge, there is no one general defense against all the inference attacks.
The reason we choose these two mechanisms is that they have been proposed to defend more diverse types of attacks compared to others.
DP is used to defend several attacks, e.g., adversarial examples~\cite{LAGHJ19}, membership inference~\cite{SS19}, model stealing~\cite{KTPPI20}, and model inversion~\cite{ZJPWLS20}.
Shejwalkar and Houmansadr~\cite{SH21} use KD to defend some inference attacks like membership inference.
Papernot et al.~\cite{PMWJS16} also introduce KD to reduce the effectiveness of adversarial examples on ML models.
Other common techniques cannot defend against all attacks simultaneously.
For instance, regularization can reduce the performance of membership inference, but regularization also leads to better model stealing and model inversion as shown in~\autoref{section:effectoverfitting}.

\subsection{Techniques}

\subsubsection{Differential Privacy (DP)}
\label{section:DP_SGD}

\begin{table*}[!t]
\centering
\setlength{\tabcolsep}{1.5pt}
\resizebox{1\textwidth}{!}{
\begin{tabular}{l|c | c | c | c | c | c | c | c | c | c | c | c}
\toprule
& \multicolumn{3}{c|}{\bf CelebA} & \multicolumn{3}{c|}{\bf FMNIST} &\multicolumn{3}{c|}{\bf STL10} &\multicolumn{3}{c}{\bf UTKFace}\\
& Original & $\epsilon_1{=}$5.139 & $\epsilon_2{=}$6.574 & Original & $\epsilon_1{=}$8.408 & $\epsilon_2{=}$9.355 & Original & $\epsilon_1{=}$8.834 & $\epsilon_2{=}$9.604 & Original & $\epsilon_1{=}$9.578 & $\epsilon_2{=}$7.762\\
\midrule
\atktu{\meminf}{\BM}{\SD} & 0.645 $\!\! \pm\!\!$ 0.005 & 0.500 $\!\! \pm\!\!$ 0.000 & 0.500 $\!\! \pm\!\!$ 0.000 & 0.559 $\!\! \pm\!\!$ 0.004 & 0.500 $\!\! \pm\!\!$ 0.000 & 0.500 $\!\! \pm\!\!$ 0.000 & 0.774 $\!\! \pm\!\!$ 0.010 & 0.501 $\!\! \pm\!\!$ 0.005 & 0.500 $\!\! \pm\!\!$ 0.000 & 0.626 $\!\! \pm\!\!$ 0.008 & 0.500 $\!\! \pm\!\!$ 0.001 & 0.499 $\!\! \pm\!\!$ 0.001\\
\atktu{\meminf}{\BM}{\PD} & 0.649 $\!\! \pm\!\!$ 0.009 & 0.500 $\!\! \pm\!\!$ 0.000 & 0.500 $\!\! \pm\!\!$ 0.000 & 0.560 $\!\! \pm\!\!$ 0.003 & 0.498 $\!\! \pm\!\!$ 0.003 & 0.500 $\!\! \pm\!\!$ 0.000 & 0.744 $\!\! \pm\!\!$ 0.028 & 0.502 $\!\! \pm\!\!$ 0.010 & 0.498 $\!\! \pm\!\!$ 0.006 & 0.621 $\!\! \pm\!\!$ 0.004 & 0.499 $\!\! \pm\!\!$ 0.007 & 0.500 $\!\! \pm\!\!$ 0.000\\
\atktu{\meminf}{\WM}{\PD} & 0.717 $\!\! \pm\!\!$ 0.004 & 0.500 $\!\! \pm\!\!$ 0.000 & 0.501 $\!\! \pm\!\!$ 0.001 & 0.580 $\!\! \pm\!\!$ 0.002 & 0.505 $\!\! \pm\!\!$ 0.001 & 0.500 $\!\! \pm\!\!$ 0.000 & 0.826 $\!\! \pm\!\!$ 0.008 & 0.511 $\!\! \pm\!\!$ 0.005 & 0.541 $\!\! \pm\!\!$ 0.002 & 0.650 $\!\! \pm\!\!$ 0.009 & 0.504 $\!\! \pm\!\!$ 0.000 & 0.505 $\!\! \pm\!\!$ 0.006\\
\atktu{\meminf}{\WM}{\SD} & 0.721 $\!\! \pm\!\!$ 0.002 & 0.500 $\!\! \pm\!\!$ 0.000 & 0.500 $\!\! \pm\!\!$ 0.000 & 0.580 $\!\! \pm\!\!$ 0.002 & 0.500 $\!\! \pm\!\!$ 0.000 & 0.500 $\!\! \pm\!\!$ 0.000 & 0.823 $\!\! \pm\!\!$ 0.005 & 0.538 $\!\! \pm\!\!$ 0.002 & 0.543 $\!\! \pm\!\!$ 0.006 & 0.660 $\!\! \pm\!\!$ 0.003 & 0.504 $\!\! \pm\!\!$ 0.000 & 0.501 $\!\! \pm\!\!$ 0.001\\
\midrule
\atktu{\modelinv}{\WM}{\SD} & 0.693 $\!\! \pm\!\!$ 0.024 & 0.640 $\!\! \pm\!\!$ 0.053 & 0.686 $\!\! \pm\!\!$ 0.053 & 0.586 $\!\! \pm\!\!$ 0.022 & 0.520 $\!\! \pm\!\!$ 0.034 & 0.570 $\!\! \pm\!\!$ 0.032 & 0.353 $\!\! \pm\!\!$ 0.008 & 0.209 $\!\! \pm\!\!$ 0.032 & 0.227 $\!\! \pm\!\!$ 0.038 & 0.912 $\!\! \pm\!\!$ 0.011 & 0.814 $\!\! \pm\!\!$ 0.020 & 0.744 $\!\! \pm\!\!$ 0.041\\
\atktu{\modelinv}{\WM}{\ND} & 0.058 $\!\! \pm\!\!$ 0.000 & 0.059 $\!\! \pm\!\!$ 0.000 & 0.059 $\!\! \pm\!\!$ 0.000 & 0.991 $\!\! \pm\!\!$ 0.000 & 0.993 $\!\! \pm\!\!$ 0.000 & 0.993 $\!\! \pm\!\!$ 0.000 & 0.201 $\!\! \pm\!\!$ 0.000 & 0.201 $\!\! \pm\!\!$ 0.000 & 0.201 $\!\! \pm\!\!$ 0.000 & 0.070 $\!\! \pm\!\!$ 0.000 & 0.071 $\!\! \pm\!\!$ 0.000 & 0.070 $\!\! \pm\!\!$ 0.000\\
\midrule
\atktu{\attrinf}{\WM}{\SD} & 0.764 $\!\! \pm\!\!$ 0.000 & 0.732 $\!\! \pm\!\!$ 0.001 & 0.701 $\!\! \pm\!\!$ 0.002 & - & - & - & - & - & - & 0.792 $\!\! \pm\!\!$ 0.002 & 0.782 $\!\! \pm\!\!$ 0.006 & 0.724 $\!\! \pm\!\!$ 0.022\\
\atktu{\attrinf}{\WM}{\PD} & 0.773 $\!\! \pm\!\!$ 0.000 & 0.732 $\!\! \pm\!\!$ 0.004 & 0.707 $\!\! \pm\!\!$ 0.002 & - & - & - & - & - & - & 0.809 $\!\! \pm\!\!$ 0.001 & 0.768 $\!\! \pm\!\!$ 0.000 & 0.740 $\!\! \pm\!\!$ 0.001\\
\midrule
\atktu{\modelsteal}{\BM}{\SD} & 0.803 $\!\! \pm\!\!$ 0.001 & 0.903 $\!\! \pm\!\!$ 0.001 & 0.896 $\!\! \pm\!\!$ 0.001 & 0.932 $\!\! \pm\!\!$ 0.001 & 0.928 $\!\! \pm\!\!$ 0.001 & 0.925 $\!\! \pm\!\!$ 0.001 & 0.663 $\!\! \pm\!\!$ 0.005 & 0.501 $\!\! \pm\!\!$ 0.010 & 0.483 $\!\! \pm\!\!$ 0.007 & 0.907 $\!\! \pm\!\!$ 0.005 & 0.845 $\!\! \pm\!\!$ 0.005 & 0.795 $\!\! \pm\!\!$ 0.006\\
\atktu{\modelsteal}{\BM}{\PD} & 0.754 $\!\! \pm\!\!$ 0.002 & 0.903 $\!\! \pm\!\!$ 0.001 & 0.895 $\!\! \pm\!\!$ 0.001 & 0.906 $\!\! \pm\!\!$ 0.004 & 0.927 $\!\! \pm\!\!$ 0.001 & 0.924 $\!\! \pm\!\!$ 0.001 & 0.525 $\!\! \pm\!\!$ 0.006 & 0.477 $\!\! \pm\!\!$ 0.006 & 0.466 $\!\! \pm\!\!$ 0.006 & 0.851 $\!\! \pm\!\!$ 0.006 & 0.838 $\!\! \pm\!\!$ 0.007 & 0.785 $\!\! \pm\!\!$ 0.004\\
\bottomrule
\end{tabular}
}
\caption{Attack performance under different threat models and datasets, on SimpleCNN, using DP-SGD.
For \meminf, \modelinv ($\tuple{\WM, \SD}$), and \attrinf, we use accuracy, for \modelinv ($\tuple{\WM, \ND}$), MSE, and for \modelsteal, agreement.}
\label{table:dpdefense}
\end{table*}

Differential Privacy (DP)~\cite{DR14,LLSY16} guarantees that any single data sample in a dataset has a limited impact on the output.

\begin{definition}[$(\epsilon,\delta)$-DP] 
\label{def:non-pure-dp}
A randomization algorithm $\AA$ satisfies $(\epsilon,\delta)$-differential privacy, with $\epsilon>0 \text{ and } 0\le \delta < 1$, if and only if for any two neighboring datasets $D$ and $D'$ that differ in one record, we have:\vspace{-0.1cm}
\begin{equation*}
\forall{T\subseteq\! \mathit{Range}(\AA)}:\; \Pr{\AA(D)\in T} \leq e^{\epsilon}\, \Pr{\AA(D')\in T}+\delta, \vspace{-0.1cm}\label{eq:npdp}
\end{equation*}
where $\mathit{Range}(\AA)$ denotes the set of all possible outputs of the algorithm $\AA$, $\delta$ can be interpreted as the probability that the mechanism fails to satisfy $\epsilon$-DP.
\end{definition}

\mypara{Gaussian Mechanism}
There are several approaches to design mechanisms satisfying $(\epsilon, \delta)$-differential privacy.
The Gaussian mechanism is arguably the most widely used one in the ML context.
Essentially, it computes a function $f$ on dataset $D$ by adding random (Gaussian) noise to $f(D)$. 
The magnitude of the noise depends on $\Delta_f$, i.e., the \textit{global sensitivity} of $f$ (also referred to as the $\ell_2$ sensitivity).
More formally, we define the $\AA$ mechanism as\vspace{-0.15cm}
\[
\AA(D) = f(D)+\calN\left(0, \Delta_f^2 \sigma^2 \mathbf{I} \right)
\]
where $\Delta_f  = \max\limits_{(D,D') : D \simeq D'} || f(D) - f(D')||_2$.

Here, $\calN (0, \Delta_f^2 \sigma^2 \mathbf{I})$ denotes a multi-dimensional random variable sampled from the normal distribution with mean $0$ and standard deviation  $\Delta_f \sigma$, and $\sigma=\sqrt{2\ln\frac{1.25}{\delta}}/\epsilon$.

\mypara{DP-SGD}
We experiment with Differentially-Private Stochastic Gradient Descent (DP-SGD)~\cite{ACGMMTZ16}, the most representative DP mechanism for protecting machine learning models.
In general, DP-SGD adds Gaussian noise to gradient $g$ during the target ML model's training process, i.e., $\Tilde{g} = g + \calN\left(0, \Delta_g^2 \sigma^2 \mathbf{I} \right)$.
Note that there is no prior knowledge to determine the influence of a single training sample on the gradient $g$; thus, the sensitivity of $g$ cannot be directly computed.
To address this problem, DP-SGD proposes to bound the $\ell_2$ norm of the gradient to $C$ by clipping $g$ to $g/\max\{1, ||g||_2 / C\}$.
This clipping ensures that if $||g||_2 \leq C$, $g$ is preserved; otherwise, it gets scaled down to be norm of $C$.
As such, the sensitivity of $g$ is bounded by $C$.

\mypara{Composition}
Note that we need to calculate the gradient multiple times when training an ML model.
Each calculation requires access to the training data and thus consumes a portion of the privacy budget.
We use the notion of \zcdp~\cite{BS16} to calculate the total privacy budget consumption.
The general idea of \zcdp is to connect $(\epsilon, \delta)$-DP to R\'enyi divergence and use the properties of R\'enyi divergence to achieve tighter composition property.
That is, for a given privacy budget $(\epsilon, \delta)$ and the number of gradient calculation $T$, \zcdp adds less Gaussian noise to the gradient than the na\"ive composition.
For instance, when $\epsilon$=1, $\delta$=1e-5, $T$=1,000, and $C$=1, the standard deviation of Gaussian noise calculated by \zcdp is 155, while that of na\"ive composition is 1,414.

\subsubsection{Knowledge Distillation (KD)}

\begin{table*}[!t]
\centering
\setlength{\tabcolsep}{5 pt}
\customTableFont
\begin{tabular}{@{}l|c | c |c | c |c | c |c | c@{}}
\toprule
& \multicolumn{2}{c|}{\bf CelebA} &\multicolumn{2}{c|}{\bf FMNIST} & \multicolumn{2}{c|}{\bf STL10} & \multicolumn{2}{c}{\bf UTKFace}  \\
& Original & Distilled & Original & Distilled & Original & Distilled & Original & Distilled\\
\midrule
\atktu{\meminf}{\BM}{\SD} & 0.595 $\pm$ 0.038 & 0.500 $\pm$ 0.000 & 0.520 $\pm$ 0.004 & 0.515 $\pm$ 0.005 & 0.682 $\pm$ 0.053 & 0.616 $\pm$ 0.069 & 0.642 $\pm$ 0.008 & 0.581 $\pm$ 0.024\\
\atktu{\meminf}{\BM}{\PD} & 0.637 $\pm$ 0.012 & 0.572 $\pm$ 0.042 & 0.530 $\pm$ 0.009 & 0.538 $\pm$ 0.008 & 0.786 $\pm$ 0.008 & 0.703 $\pm$ 0.005 & 0.657 $\pm$ 0.009 & 0.596 $\pm$ 0.001\\
\atktu{\meminf}{\WM}{\PD} & 0.698 $\pm$ 0.005 & 0.691 $\pm$ 0.002 & 0.555 $\pm$ 0.013 & 0.568 $\pm$ 0.001 & 0.806 $\pm$ 0.015 & 0.773 $\pm$ 0.006 & 0.677 $\pm$ 0.000 & 0.611 $\pm$ 0.002\\
\atktu{\meminf}{\WM}{\SD} & 0.638 $\pm$ 0.038 & 0.559 $\pm$ 0.059 & 0.539 $\pm$ 0.014 & 0.530 $\pm$ 0.004 & 0.799 $\pm$ 0.050 & 0.677 $\pm$ 0.073 & 0.642 $\pm$ 0.017 & 0.620 $\pm$ 0.024\\
\midrule
\atktu{\modelinv}{\WM}{\SD} & 0.648 $\pm$ 0.038 & 0.650 $\pm$ 0.030 & 0.573 $\pm$ 0.021 & 0.447 $\pm$ 0.037 & 0.203 $\pm$ 0.020 & 0.244 $\pm$ 0.031 & 0.824 $\pm$ 0.021 & 0.815 $\pm$ 0.036\\
\atktu{\modelinv}{\WM}{\ND} & 0.058 $\pm$ 0.000 & 0.058 $\pm$ 0.000 & 0.993 $\pm$ 0.000 & 0.992 $\pm$ 0.000 & 0.201 $\pm$ 0.000 & 0.201 $\pm$ 0.000 & 0.070 $\pm$ 0.000 & 0.070 $\pm$ 0.000\\
\midrule
\atktu{\attrinf}{\WM}{\SD} & 0.665 $\pm$ 0.005 & 0.669 $\pm$ 0.019 & - & - & - & - & 0.540 $\pm$ 0.006 & 0.554 $\pm$ 0.030\\
\atktu{\attrinf}{\WM}{\PD} & 0.610 $\pm$ 0.018 & 0.660 $\pm$ 0.004 & - & - & - & - & 0.549 $\pm$ 0.001 & 0.584 $\pm$ 0.004\\
\midrule
\atktu{\modelsteal}{\BM}{\SD} & 0.820 $\pm$ 0.001 & 0.788 $\pm$ 0.003 & 0.932 $\pm$ 0.000 & 0.940 $\pm$ 0.001 & 0.589 $\pm$ 0.006 & 0.618 $\pm$ 0.003 & 0.927 $\pm$ 0.007 & 0.918 $\pm$ 0.013\\
\atktu{\modelsteal}{\BM}{\PD} & 0.756 $\pm$ 0.005 & 0.741 $\pm$ 0.003 & 0.902 $\pm$ 0.001 & 0.914 $\pm$ 0.001 & 0.478 $\pm$ 0.006 & 0.510 $\pm$ 0.003 & 0.826 $\pm$ 0.010 & 0.836 $\pm$ 0.018\\
\bottomrule
\end{tabular}
\caption{Attack performance under different threat models and datasets, on VGG19, using Knowledge Distillation (KD).
For \meminf, \modelinv ($\tuple{\WM, \SD}$), and \attrinf, we use accuracy, for \modelinv ($\tuple{\WM, \ND}$), MSE, and for \modelsteal, agreement.}
\label{table:distdefense}
\end{table*}

Another defense mechanism we consider is Knowledge Distillation (KD)~\cite{HVD15,SH21}.
Generally, KD is proposed to transfer the generalization ability (knowledge) from a larger model (original model) to a smaller model (distilled model) without utility degradation.
Once the distilled model is trained, it can replace the original model in many scenarios as it is more computationally efficient and less dependent on resources.

A simple way to transfer the knowledge from the original model to the distilled model is to use the posteriors generated by the original model as a ``soft label'' to guide the training of the distilled model.
Compared to the original labels (one-hot), the posteriors have higher entropy.
It contains more information for each training sample and has less variance for the gradient among different training samples, which can speed up the training process of the distilled model~\cite{HVD15}. 
To train the distilled model, we combine two loss terms, i.e., the soft target loss and the hard target loss.
The first one is the Kullback-Leibler divergence loss between the output of the original model and the distilled model.
The second one is the cross-entropy loss between the original label and the output of the distilled model.
As suggested by Hinton et al.~\cite{HVD15}, we use a higher temperature value in the softmax function of the first loss for better performance.

KD transfers knowledge from the original model to a distilled model.
Compared to the original model, the distilled model has a lower capacity.
Intuitively, it should remember less information of the original model with respect to both its training dataset and parameters. 
Thus, we believe KD can serve as a general defense for inference attacks.
Papernot et al.~\cite{PMWJS16} show that KD can reduce the risks of adversarial examples against machine learning models.
Shejwalkar and Houmansadr~\cite{SH21} also show that KD can mitigate membership inference attacks.
Here, we take a broader view investigating whether or not KD is effective to defend against other inference attacks.

\subsection{Experimental Setup}

Both DP-SGD and KD are applied in the training process of target models.
Due to space limitations, we only apply DP-SGD to SimpleCNN and KD to VGG19.

\mypara{DP-SGD Target Model}
We use the Opacus library\footnote{\url{https://github.com/pytorch/opacus}} to implement DP-SGD.
This library allows a user to configure the clip bound $C$, the standard deviation of the Gaussian noise $\sigma$, and the failure probability $\delta$, then the library can automatically calculate the total privacy budget $\epsilon$ using \zcdp.
A larger number of epochs implies higher $\epsilon$.
Our target model is trained for 300 epochs; thus, we fix $\delta$=1e-5 and choose two sets of $C$ and $\sigma$ such that $\epsilon$ is smaller than 10.
We list these settings in the second row of~\autoref{table:dpdefense}.
All the other hyperparameters are the same as presented in \autoref{section:target_model}.

\mypara{Distillation Target Model}
We distill the model knowledge of VGG19 (16 convolution layers and 3 fully connected layers) to a smaller model, i.e., VGG11~\cite{SZ15} (8 convolution layers and 3 fully connected layers).
We use Kullback-Leibler divergence as the soft target loss with the temperature setting to 20.
For the hard target loss, we use cross-entropy.
We set $\alpha$ to 0.7 for the ratio of the soft target loss.
Other settings are the same as the target model's training phase in \autoref{section:target_model}.

\subsection{Results}

\begin{table}[!t]
\centering
\customTableFont
\setlength{\tabcolsep}{5 pt}
\begin{tabular}{@{}llc c c c@{}}\toprule
{\bf Experiment} & {\bf Model} & {\bf CelebA} & \hspace{-0.1cm}{\bf FMNIST} & \hspace{-0.1cm}{\bf STL10} & \hspace{-0.1cm}{\bf UTKFace}\\
\midrule
{\bf DP-SGD} ($\epsilon_1$) & SimpleCNN & 0.654 & 0.830 & 0.347 & 0.698\\
{\bf DP-SGD} ($\epsilon_2$) & SimpleCNN & 0.675 & 0.836 & 0.313 & 0.680\\
{\bf KD} & VGG19 & 0.713 & 0.919 & 0.588 & 0.823\\
\midrule
{\bf No Defense} &  SimpleCNN & 0.706 & 0.903 & 0.516 & 0.818\\
{\bf No Defense} & VGG19 & 0.733 & 0.905 & 0.587 & 0.834\\
\bottomrule
\end{tabular}
\caption{Accuracy of target models protected by DP-SGD and KD.}
\label{table:defense_utility}
\end{table}

\mypara{DP-SGD}
\autoref{table:dpdefense} reports the performance of inference attacks against target models protected by DP-SGD.
For \meminf, DP-SGD is effective in almost all cases.
For instance, for \atktu{\meminf}{\WM}{\SD} on the CelebA dataset, the accuracy drops from 0.721 to 0.500, which is a random guess.
This is expected as DP, by definition, can mitigate membership inference.
For \modelinv and \attrinf, DP-SGD can only reduce the attack accuracy to a small extent.
However, the MSE loss for \atktu{\modelinv}{\WM}{\ND} remains stable.

DP-SGD indeed reduces the risks of model stealing on STL10 and UTKFace under different threat models.
For instance, for \atktu{\modelsteal}{\BM}{\SD} on STL10, the agreement for the two different $\epsilon$s are 0.501 and 0.483, while the agreement for the original model is 0.663.
Meanwhile, DP-SGD does not influence model stealing on FMNIST.
Interestingly, it actually enhances the performance of model stealing on CelebA.
Overall, DP-SGD can effectively defend against membership inference attacks, but not the others. 

\mypara{KD}
In~\autoref{table:distdefense}, we report the effectiveness of KD as a general defense mechanism.
We do not observe any significant decrease in attack performance for model inversion, attribute inference, and model stealing on original vs.~the distilled models.
Specifically, the attack performance difference, in most cases, is less than 5\%.
In certain cases, KD is effective against membership inference, but to a lesser extent compared to DP-SGD.
For instance, the accuracy of \atktu{\meminf}{\BM}{\SD} on the STL10 dataset drops from 0.682 to 0.616.

\mypara{Utility and Defense Effectiveness Trade-off}
We observe that both DP-SGD and KD can defend some of the inference attacks. 
However, it comes at the cost of utility dropping (see~\autoref{table:defense_utility}).
Compared to DP-SGD, KD preserves the target model's utility better.
For instance, on the STL10 dataset, the target testing accuracy drops from 0.818 to 0.698 ($\epsilon_1$) and 0.680 ($\epsilon_2$) for DP-SGD, while the corresponding performance only drops from 0.834 to 0.823 for KD.
Meanwhile, DP-SGD has better defense performance than KD.
In particular, for \atktu{\meminf}{\WM}{\SD} on the UTKFace dataset, SimpleCNN defended by DP-SGD reduces the attack accuracy significantly from 0.660 to 0.504 ($\epsilon_1$) and 0.501 ($\epsilon_2$), respectively.
The VGG19 model defended by KD only reduces the attack accuracy to a smaller extends from 0.642 to 0.621.
Also, as discussed above, both DP-SGD and KD are not general defenses against all the inference attacks, which inspires future research to better defend different inference attacks while maintaining the target model's utility.

\section{Related Work}
\label{section:relatedwork}

We now review relevant related work on inference attacks and defenses, as well as software dedicated to evaluating them.

\mypara{Membership Inference Attacks}
Shokri et al.~\cite{SSSS17} propose the first membership inference attack against black-box ML models: they train multiple shadow models to simulate the target model and use multiple attack models to conduct the inference.
Salem et al.~\cite{SZHBFB19} later relax several key assumptions from~\cite{SSSS17}; namely, using multiple shadow models, the knowledge of the target model structure, and having a dataset from the same distribution as the target model’s.
Yeom et al.~\cite{YGFJ18} assume that the adversary knows the target model's training dataset's distribution and size, and they collude with the training algorithm.
Both~\cite{SZHBFB19} and~\cite{YGFJ18} are close in performance to Shokri et al.'s attacks~\cite{SSSS17}. 
In this paper, we implement the attack proposed by Salem et al.~\cite{SZHBFB19}, i.e., one shadow model, one attack model, and a shadow dataset.
More recently, researchers have studied membership inference in other settings, including natural language processing~\cite{SS19,CTWJHLRBSEOR20}, generative models~\cite{CYZF20,HMDC19,HHB19}, recommender systems~\cite{ZRWRCHZ21}, and federated learning~\cite{MSCS19,NSH19}.
Also, Song and Mittal have performed a systematic evaluation on membership inference~\cite{SM21}.
Previous work~\cite{SSSS17,SZHBFB19} also shows that overfitting is the major factor causing membership inference.
To the best of our knowledge, however, no one has investigated other factors studied in our paper, such as the influence of dataset complexity or the relationship among different inference attacks. 

\mypara{Attribute Inference}
Prior research~\cite{AMSVVF15,GWYGB18} has studied macro-level attribute inference attacks against ML models, whereby the adversary aims to infer some general properties of the training dataset.
Melis et al.~\cite{MSCS19} propose the first sample-level attribute inference attack against federated machine learning systems.
Song and Shmatikov~\cite{SS20} reveal that the risks of attribute inference are caused by the intrinsic overlearning characteristics of machine learning models.

\mypara{Model Inversion}
Model inversion is first proposed by Fredrikson et al.~\cite{FLJLPR14} in the setting of drug dose classification.
Later, they extend model inversion to general ML settings relying on back-propagation over a target ML model's parameters~\cite{FJR15}.
More recently, Zhang et al.~\cite{ZJPWLS20} develop a more advanced attack aiming to synthesize the training dataset relying on GANs.
Finally, Carlini et al.~\cite{CLEKS19} show that model inversion can be effectively performed against natural language processing models as well. 

\mypara{Model Stealing}
Tram{\`e}r et al.~\cite{TZJRR16} propose the first model stealing attack against black-box machine learning API.
Orekondy et al.~\cite{OSF19} develop a reinforcement learning-based framework to optimize both query time and effectiveness.
Also, Wang and Gong~\cite{WG18} and Oh et al.~\cite{OASF18} show that hyperparameters of a target model can be inferred as well.

\mypara{Defense Mechanisms}
A few defense mechanisms have been proposed to mitigate membership inference attacks~\cite{NSH18,SZHBFB19,JSBZG19}.
However, these defenses are specifically designed for membership inference and cannot mitigate other inference attacks.
For instance, Salem et al.~\cite{SZHBFB19} propose to reduce overfitting of the target model as a defense; however, as we show in our analysis (see \autoref{section:effectoverfitting}), reducing overfitting will improve the performance of model stealing.

Differential Privacy (DP)~\cite{DR14,LLSY16} guarantees that any single data sample in a dataset has a limited impact on the output of an algorithm.
As such, it is an effective defense mechanism against inference attacks.
Abadi et al.~\cite{ACGMMTZ16} introduce DP-SGD, which adds Gaussian noise to the gradients of the target model during the training process.
Another DP method for protecting the privacy of ML models is PATE~\cite{PAEGT17}: a set of teacher models is trained on a private dataset, which is used to label a public dataset in a differentially private manner.
The final public dataset is then used to train a student model.
Recently, Nasr et al.~\cite{NSTPC21} instantiate a number of attacks against ML to evaluate the effectiveness of DP defenses and, in particular, how tight are theoretical DP bounds.

Another defense mechanism, as mentioned, is Knowledge Distillation (KD)~\cite{HVD15}.
Papernot et al.~\cite{PMWJS16} propose a defensive distillation mechanism to effectively reduce the risks for target models with respect to adversarial examples.
Shejwalkar and Houmansadr~\cite{SH21} reveal that distillation can reduce the gap between the posteriors of members and non-members, thus protecting membership privacy. 
In our experiments, we show that distillation is indeed effective against certain target models supported by \systemName; however, it cannot defend against other types of inference attacks.

\mypara{Risk Assessment Tools}
Finally, researchers have recently developed a number of software tools to measure the potential security/privacy risks of ML models.
Ling et al.~\cite{LJZWWLW19} propose DEEPSEC, a security analysis system to evaluate different adversarial example attacks and defenses. 
Another system for adversarial examples is CleverHans~\cite{PFCGFKXSBRMBHZJLSGUGDBHRLM18}.
Pang et al.~\cite{PZGXJCW20} introduce TROJANZOO, which focuses on backdoor attacks.

Closer to our work is ML Privacy Meter~\cite{MS20}, which jointly considers membership inference attacks in both black-box and white-box settings. 
Unlike ML Privacy Meter, which focuses on membership inference only, \systemName considers four types of inference attacks simultaneously.
In addition, we rely on \systemName to perform a comprehensive analysis for all these inference attacks.

\section{Conclusion}
\label{section:conclusion}

In this paper, we performed the first holistic analysis of privacy risks caused by inference attacks against machine learning models.
We established a taxonomy of threat models for four types of inference attacks, including membership inference, model inversion, attribute inference, and model stealing.
We conducted an extensive measurement study, over five model architectures, and four datasets, of both attacks and defenses.
Among other things, we found that the complexity of the training dataset plays an important role in the attack's performance, while the effectiveness of model stealing and membership inference attacks are negatively correlated.
We also showed that defenses such as DP-SGD and KD could only hope to mitigate {\em some} of the inference attacks.

We integrated all the attacks and defenses into a re-usable, modular software called \systemName, which can be used in various scenarios.
For instance, an ML model owner can use \systemName to evaluate the model's inference risks before deploying it in the real world.
We are also confident that \systemName will serve as a benchmark tool to facilitate future research on inference attacks and defenses.

Currently, \systemName concentrates on image classification models, as image classification is the most popular ML application.
Researchers have demonstrated that inference attacks can be successfully launched against other types of ML models, such as language models~\cite{SS19,SR20}, generative models~\cite{HMDC19,CYZF20}, and graph-based models~\cite{HJBGZ21}, as well as other training paradigms, such as federated learning~\cite{MSCS19}.
We plan to extend \systemName to support a broader range of ML application scenarios.
In addition, we will explore other general defense mechanisms, such as training target models with noisy data or GAN-generated data.

Finally, while \systemName is designed for inference attacks, we plan to integrate tools~\cite{PFCGFKXSBRMBHZJLSGUGDBHRLM18,LJZWWLW19,PZGXJCW20} geared to evaluate risks aimed to jeopardize models' functionality, e.g., adversarial examples, data poisoning, etc., thus providing a one-stop-shop toward enabling secure and trustworthy AI.

\mypara{Acknowledgments}
The authors wish to thank Luca Melis and Jamie Hayes for valuable discussions and feedback.
This work is partially funded by the Helmholtz Association within the project ``Trustworthy Federated Data Analytics'' (TFDA) (funding number ZT-I-OO1 4).

\balance
\bibliographystyle{plain}
\bibliography{normal_generated_py3}

\appendix
\section{Additional Experimental Results}
\label{section:appendix}

In this appendix, we report plots for additional experiments as mentioned throughout the paper.

\begin{figure}[!bp]
\centering
\includegraphics[width=0.8\columnwidth]{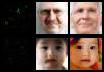}
\label{figure:modinvwsm}
\caption{Visualization of model inversion (AlexNet trained on UTKFace).
The left column depicts two samples reconstructed using~\cite{FJR15},
the middle one using~\cite{ZJPWLS20},
while the right column reports two samples from the target model's training dataset.
The left column images are normalized, black indicating pixels' value in reconstructed images are close to 0.
Note that similar results are shown by Zhang et al.~\cite{ZJPWLS20}.
}
\label{figure:ccs15reverse}
\end{figure}

\begin{figure*}[!ht]
\centering
\includegraphics[width=1.9\columnwidth]{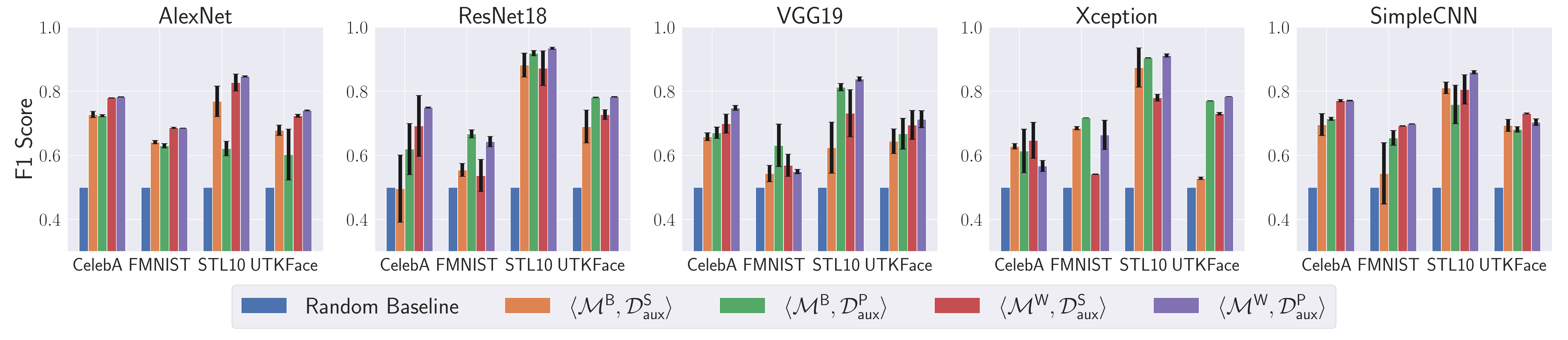}
\caption{F1 score of membership inference attacks (\meminf) under different threat models, datasets, and target model architectures.}
\label{figure:meminf_f1}
\end{figure*} 

\begin{figure*}[!ht]
\centering
\includegraphics[width=1.9\columnwidth]{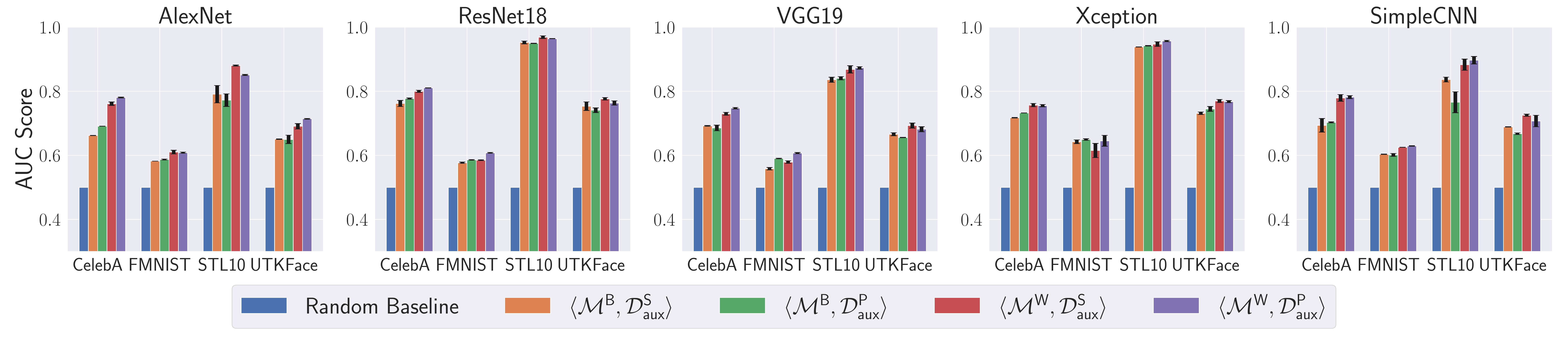}
\caption{AUC score of membership inference attacks (\meminf) under different threat models, datasets, and target model architectures.}
\label{figure:meminf_auc}
\end{figure*} 

\begin{figure*}[!ht]
\centering
\includegraphics[width=1.9\columnwidth]{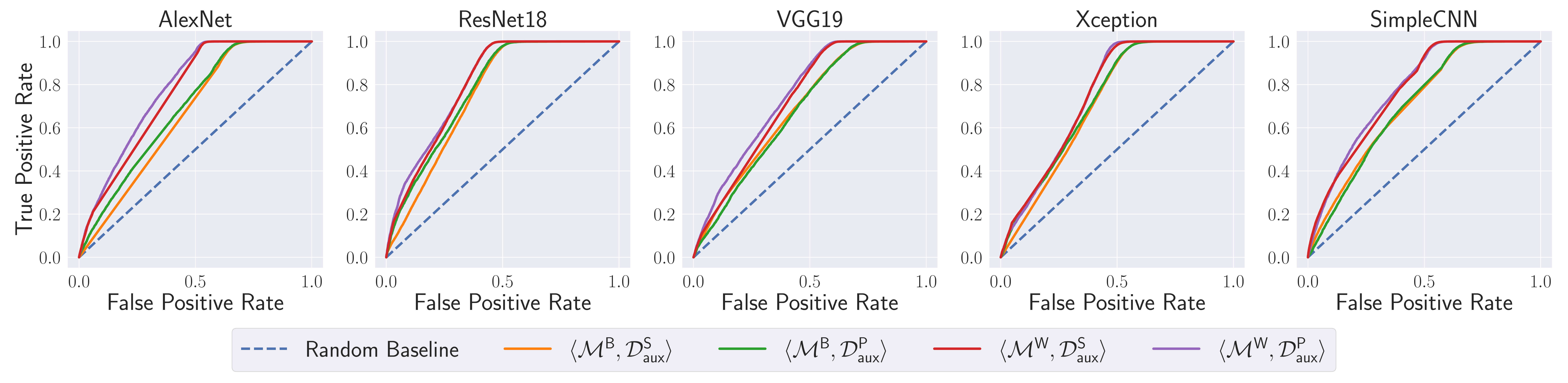}
\caption{ROC curve of membership inference attacks (\meminf) under different threat models on CelebA.}
\label{figure:roc_celeba}
\end{figure*}

\begin{figure*}[!ht]
\centering
\includegraphics[width=1.9\columnwidth]{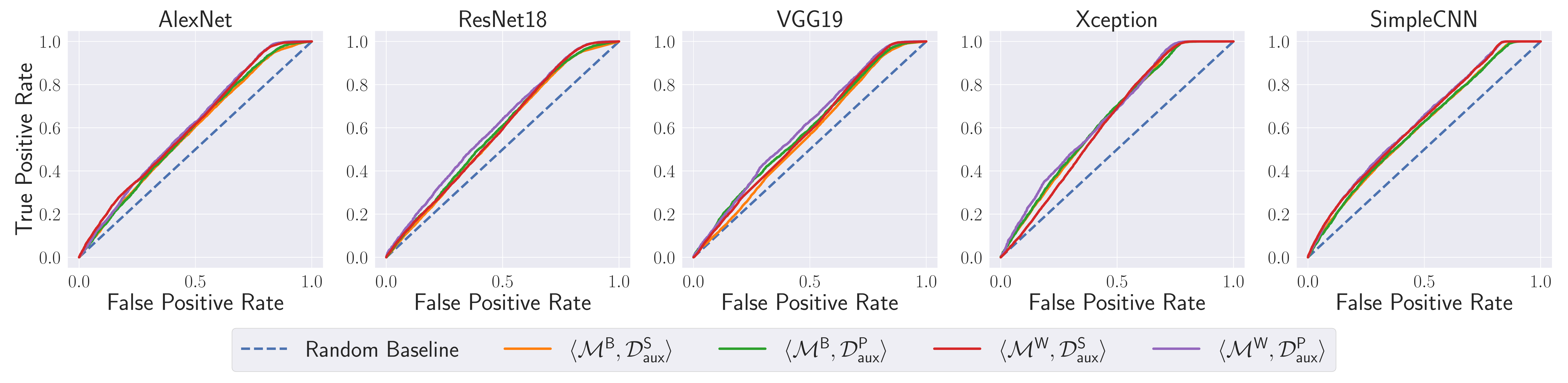}
\caption{ROC curve of membership inference attacks (\meminf) under different threat models on FMNIST.}
\label{figure:roc_FMNIST}
\end{figure*}

\begin{figure*}[!ht]
\centering
\includegraphics[width=1.9\columnwidth]{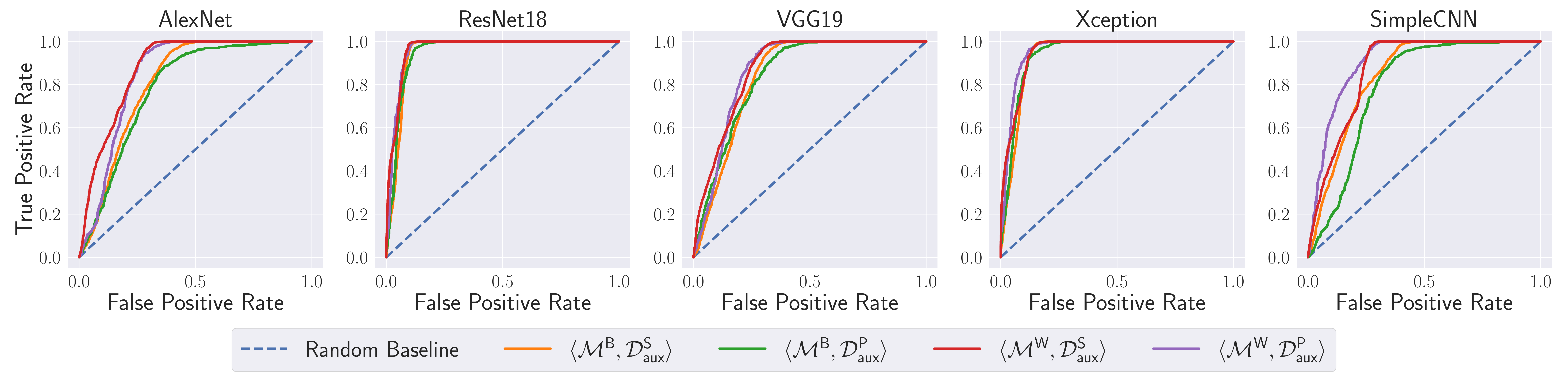}
\caption{ROC curve of membership inference attacks (\meminf) under different threat models on STL10.}
\label{figure:roc_stl10}
\end{figure*}

\begin{figure*}[!ht]
\centering
\includegraphics[width=1.9\columnwidth]{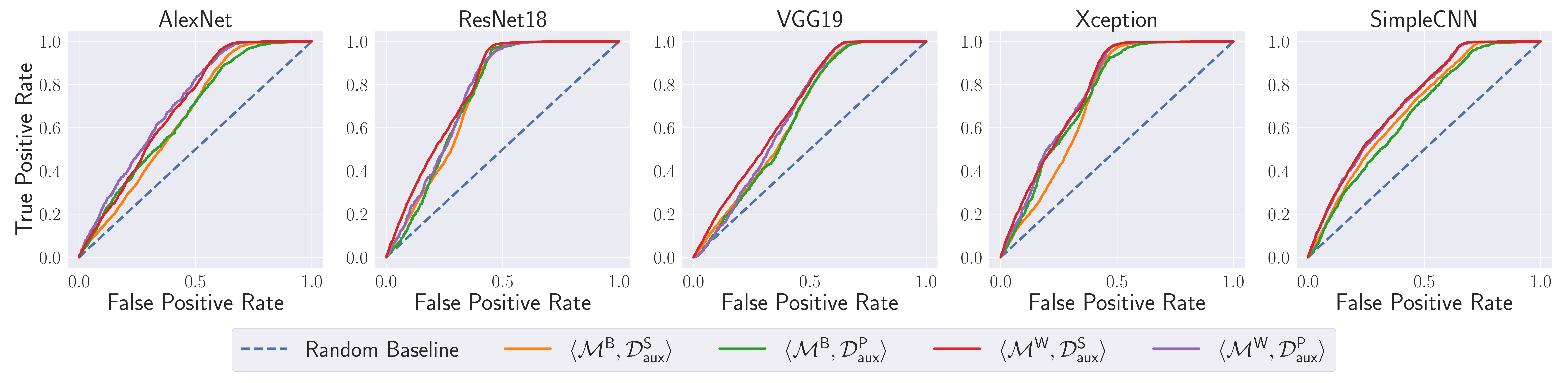}
\caption{ROC curve of membership inference attacks (\meminf) under different threat models on UTKFace.}
\label{figure:roc_UTKFace}
\end{figure*}

\begin{figure*}[!ht]
\centering
\includegraphics[width=1.9\columnwidth]{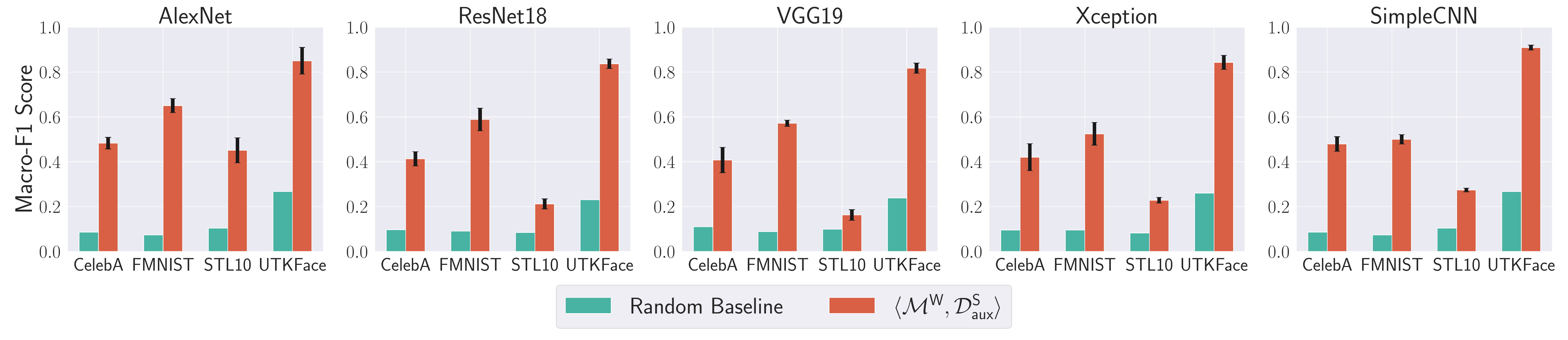}
\caption{Macro-F1 score of model inversion attacks (\modelinv) under different datasets and target model architectures.}
\label{figure:modinv_f1}
\end{figure*} 

\begin{figure*}[!ht]
\centering
\includegraphics[width=1.9\columnwidth]{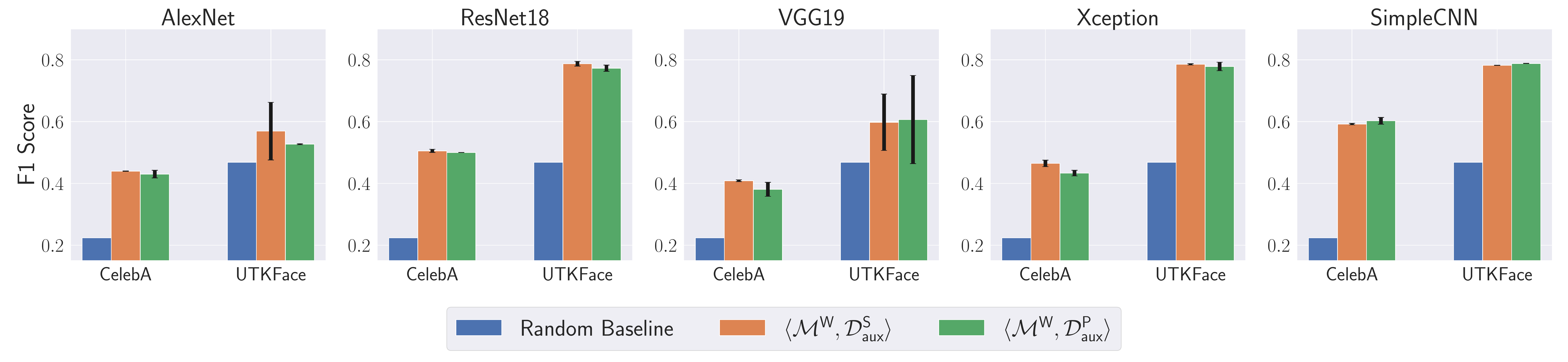}
\caption{F1 score of attribute inference attacks (\attrinf) under different threat models, datasets, and target model architectures. 
Note that we report F1 score for UTKFace and macro-F1 score for CelebA.}
\label{figure:attrinf_f1}
\end{figure*}

\begin{figure*}[!ht]
\centering
\includegraphics[width=1.9\columnwidth]{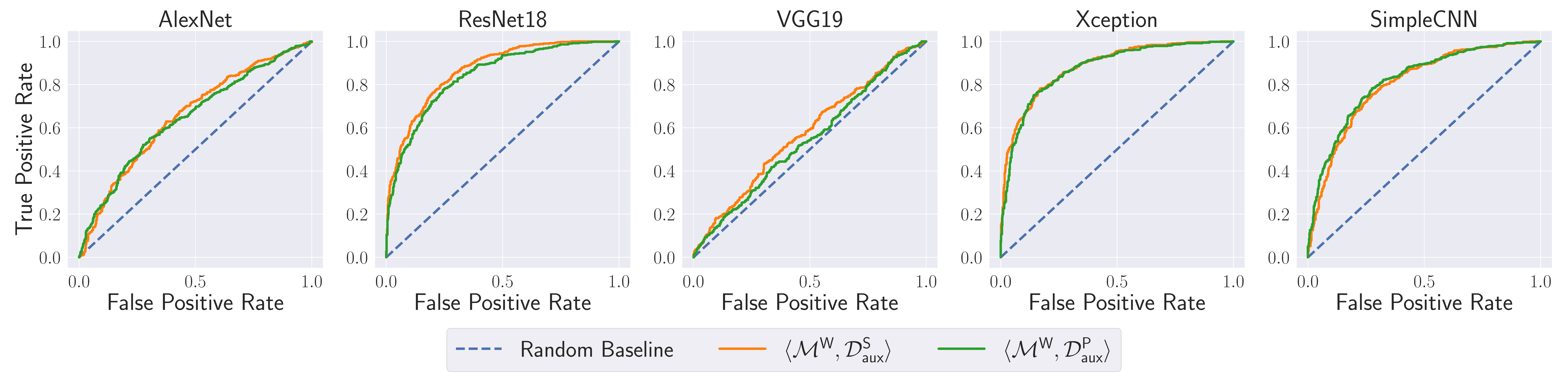}
\caption{ROC curve of attribute inference attacks (\attrinf) on UTKFace under different threat models and target model architectures.}
\label{figure:roc_utkface_attrinf}
\end{figure*}

\begin{figure*}[!ht]
\centering
\includegraphics[width=1.9\columnwidth]{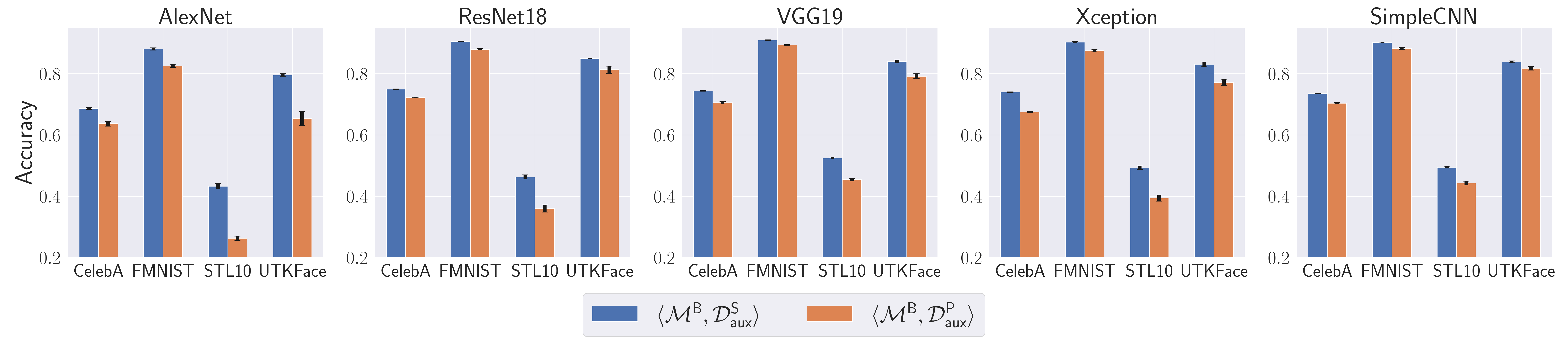}
\caption{Accuracy of model stealing attacks (\modelsteal) under different threat models, datasets, and target model architectures.}
\label{figure:modsteal_acc}
\end{figure*} 

\begin{figure*}[!ht]
\centering
\includegraphics[width=1.5\columnwidth]{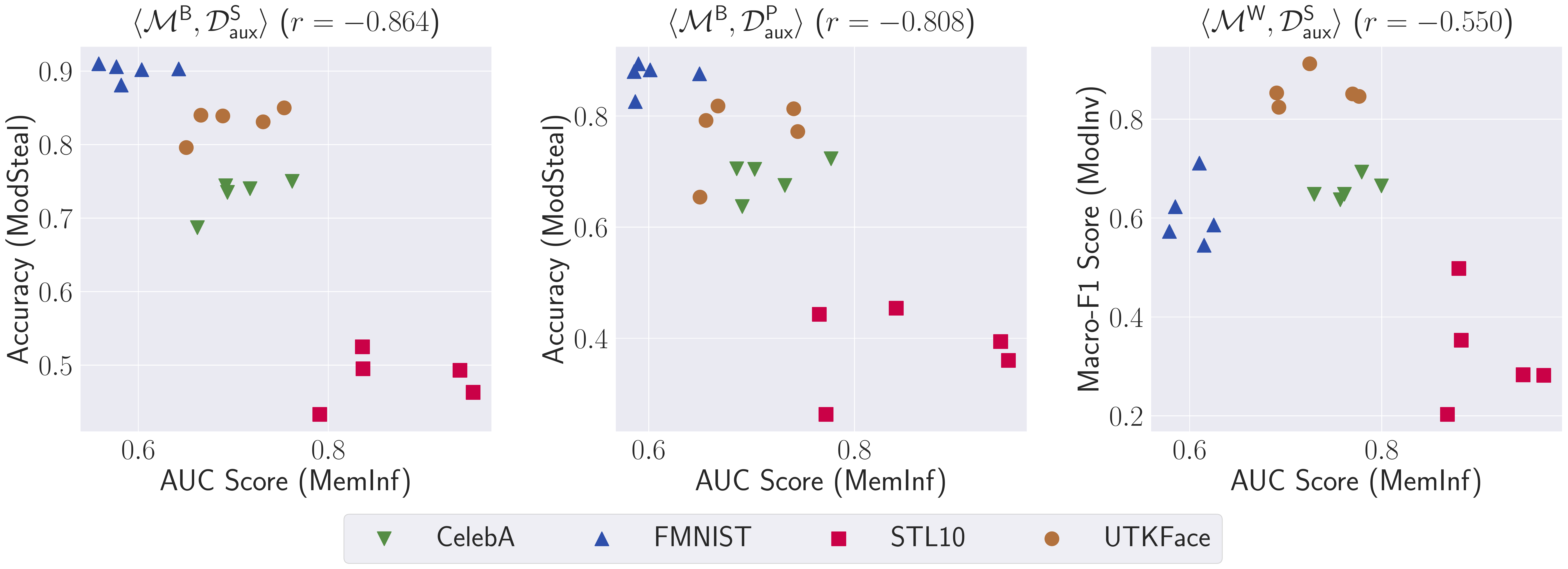}
\caption{The relation between different attacks under the same threat model.
For \meminf, we use AUC score, for \modelinv, macro-F1, and for \modelsteal, accuracy.}
\label{figure:correlation_auc}
\end{figure*}

\end{document}